\documentclass[prd,aps,floatfix,nofootinbib,preprint ,tightenlines,superscriptaddress]{revtex4}
\usepackage{dcolumn}
\usepackage{amsmath}
\usepackage{latexsym}
\usepackage{graphicx}
\usepackage{bm}

\def\svev#1{\left\langle #1\right\rangle}       

\def\Tr{{\rm Tr}\,}
\def\Re{{\rm Re\,}}

\def\det{{\rm det}}

\def\ie{{\it i.e.,\ }}
\newcommand{\VEV}[1]{\left\langle #1\right\rangle}

\newcommand{\bra}[1]{\left\langle #1 \right|}
\newcommand{\ket}[1]{\left| #1 \right\rangle}

\newcommand{\bee}{\begin{equation}}
\newcommand{\ee}{\end{equation}}
\newcommand{\beea}{\begin{eqnarray}}
\newcommand{\eea}{\end{eqnarray}}

\begin{document}
\title{Lattice methods for students at a formal TASI}
\author{Thomas DeGrand}
\affiliation{
Department of Physics, University of Colorado,
        Boulder, CO 80309 USA}
\email{thomas.degrand@colorado.edu}

\date{\today}

\begin{abstract}
These lectures about lattice field theory were written for, and given at, TASI 2019, ``The many 
dimensions of quantum field theory.''
The students at this TASI were mostly interested in formal things, and so these are 
slightly unusual lattice lectures:
I wanted to give the physical motivation behind lattice calculations
rather than describe all the technical details.
A quick outline: (1) The really big picture: lattice basics, lattice confinement, 
getting rid of the lattice.
(2) A walk through the parts of a lattice calculation -- an overview,
to show what's involved.
(3) Chiral fermions on the lattice. (This part might be interesting to lattice people.)
(4) Case studies: the three dimensional Ising model, and QCD.
\end{abstract}
\maketitle

\section{Introduction}

What is the lattice? Replace continuous space or space-time with a grid, 
 put all your dynamical variables (fields) on the grid. You can do this for a
Hamiltonian
\bee
H= \int d^D x {\cal H} \rightarrow \sum_i H_i
\ee
or for a path integral
\bee
Z= \int [d\phi] e^{-S} \rightarrow \int \prod_i d\phi_i e^{-S} .
\ee
The grid could be a set of space-time points, or space points (keeping time still continuous). The grid
could be anything you want, regular or irregular (even random). I'll assume we have a simple 
hypercubic  grid
and call the spacing between the points on the grid the ``lattice spacing'' $a$.

Why do this? Sometimes, the physics you want to study actually lives on a grid. (Presumably, that is more appropriate
for students at the Boulder School, next month). More often, your system is complicated, possibly strongly interacting,
and you hope that by thinning out the number of degrees of freedom, you can make progress with it.
  You want to preserve all your ``internal'' (global or gauge) symmetries, 
but you are willing to sacrifice the space-time ones.
You are willing to live with a UV cutoff (the lattice spacing $a$), an IR cutoff (if the number of points
in the grid is finite), discretization effects (you replaced derivatives by differences), or, more often,
 you think you can postpone dealing with these issues until later.
What do you get?
\begin{itemize}
\item A UV cutoff with no connection to perturbation theory
\item
  Nonperturbative insight (sometimes). For example,
 putting gauge theories on a lattice led Wilson to a description of confinement.
\item
  Accessibility to numerical simulation (sometimes)
\end{itemize}

This is an old subject. It's been part of particle physics for 45 years and a part of condensed matter physics
 for much longer.
Probably none of you  work on it. And there are perfectly good books to read, if you want to do it.
So  why should you listen to lectures about it?

I think that to best reason to listen is that you might find lattice techniques useful in some problem down the road.
If you don't want to write your own Monte Carlo program (and I can't blame you for this), you probably
will want to persuade some lattice expert to help you -- or do it for you. It might be good to know
what you can expect to get out of a lattice calculation, as well as how lattice people approach their problems.

Along the way, I can tell you a bit about QCD. This is the system most lattice people study. Lattice QCD
is probably the most successful program in particle theory (in terms of actual numbers to compare
to experiment) since the formulation of the Standard Model.
But lattice is not just for QCD. People use it for all sorts of systems.

So I will start with our creation myth: how to put gauge fields on the lattice, how the Wilson loop is
an indicator of confinement, and (most important of all) how we take the lattice spacing away.

Lattice calculations are as technical as anything else which is discussed at a Tasi.
 I am going to try to avoid
technical details as much as I can, with the goal of giving you motivation to read more and to understand
the language spoken by lattice people. But you have to know the limitations of any technique. 
Lecture 2 will be a long walk through all the parts
of a lattice simulation, with maybe more than you might want to know.
In Lecture 3
 I will describe
 how to deal with chiral fermions on the lattice (this has connections to the topological insulator game),
Then I want to do case studies (with pictures).
I'll look at a lattice calculation of critical indices in the three dimensional Ising model.
Finally, QCD.  I'll do something very un-lattice here: I'll try to tell you an organizational story.
All confining systems look very similar.

I know of three good recent books about lattice techniques. They are all very QCD-centric:
They are DeGrand and DeTar \cite{DeGrand:2006zz},
Gattringer and Lang \cite{Gattringer:2010zz},
and
Knechtli, G\"unther and Peardon \cite{Knechtli:2017sna}.
There is also a recent preprint by Hanada \cite{Hanada:2018fnp} which is absolutely not
QCD-centric. It has snippets of code (which is nice; modern QCD codes available for download are 
a bit daunting)
and many opinions (not all of which I agree with).

\section{The creation myth of lattice gauge theory}
\subsection{Path integral formulation}
At zeroth order, formulating a lattice action or Hamiltonian for a system with global symmetries
is no big deal. Just put your fields on sites, and replace derivatives by finite differences.
The usual definition of a global symmetry  transformation (I am assuming that $V^\dagger V=1$)
\bee
\phi(x) \rightarrow V \phi(x); \qquad  \phi^\dagger(x) \rightarrow  \phi^\dagger(x) V^\dagger
\ee
naturally discretizes  by just attaching an index to $x$ ($\phi(x_i)$). 
Path integral expectation values encode the symmetry because we can perform the global
 transformation under the integral,
rotating the variables we integrate over, and pushing the variable change onto the observable.
That's about all there is to say, now we can go do physics.

However, a local gauge transformation is
\bee
\phi(x) \rightarrow V(x) \phi(x); \qquad  \phi^\dagger(x) \rightarrow  \phi^\dagger(x) V^\dagger(x)
\label{eq:lgt}
\ee
and if the action was invariant under this transformation, then an expectation value 
can be transformed to
\bee
\svev{\phi(x')^\dagger \phi(x)} = \svev{\phi^\dagger(x') V^\dagger(x') V(x) \phi(x)}.
\ee
This has to vanish because $V^\dagger(x') V(x)$ could be anything. This is
just a complicated way of saying that the
 correlator is not gauge invariant, and expectation values of gauge non-invariant objects vanish.
 Dealing with this issue is the same on the lattice as it is in the continuum,
local gauge invariance requires the presence of additional dynamical variables. 
Our continuum-based textbooks tell us that a
particle traversing a contour in space from a point $x$ to a point $y$ picks up a connection factor
\bee
\phi_s(y) \rightarrow P(\exp \ ig \int_s dx_\mu A_\mu) \phi(x)
 \equiv U(y,x)\phi(x) \label{2.3}
\ee
where
$U$ is an element of the symmetry group and
$P$ is the path-ordering factor.
 Under a gauge transformation, when the matter field is locally rotated,
\bee
\phi(x)\rightarrow V(x) \phi(x),
\ee
 $U(s)$ is rotated at each end
\bee
U(s) \rightarrow V(x(s))U(s)V(x(0))^\dagger. \label{eq:2.4}
\ee
so that objects like $\phi(x(s))^\dagger U(s) \phi(x(0))$ are gauge invariant.

We have to get $U$-like variables into the functional integral.
The minimum length path connects adjacent sites on the lattice, and so
 we choose (following Wilson \cite{Wilson:1974sk})
to work with
 fundamental degrees of freedom which are elements of the gauge group, living
on the links of the lattice, connecting $x$ and $x+ \mu$:
$U_\mu(x)$, with $U_{-\mu}(x+\mu) = U_\mu(x)^\dagger$. 
In general, $U$ is just a group element, but we
 can also write it as
 \bee
U_\mu(n)= \exp (igaT^aA^a_\mu(n))  \label{eq:2.5}
\ee
for a typical continuous symmetry transformation
($g$ is the coupling, $a$ the lattice spacing, $A_\mu$ the vector potential,
and $T^a$  a group generator). 
Under a gauge transformation, link variables transform as a special case of Eq.~\ref{eq:2.4},
\bee
U_\mu (x) \rightarrow V(x) U_\mu (x) V(x+ \hat \mu)^\dagger.  \label{eq:2.10}
\ee
(Like most lattice people, I have set $a=1$.)
Gauge invariant operators we can use as order parameters will be
 matter fields connected by  oriented ``strings" of U's like Fig. 1a,
\bee
\bar \psi(x) U _\mu(x)U_\mu(x+\hat \mu)\ldots  \psi (x_2) , \label{2.13}
\ee
or closed,  oriented loops of U's like Fig. 1b,
\bee
{\rm Tr} \ldots U _\mu(x)U_\mu(x+\hat \mu)\ldots \rightarrow
{\rm Tr} \ldots U_\mu(x)V^\dagger (x+ \hat \mu)V(x+ \hat \mu)
U_\mu(x+\hat \mu)\ldots  .\label{eq:2.12}
\ee

The functional integration measure for the links is the so called ``Haar measure,''
invariant integration over the group elements. 

So much for the definition of variables.
An action for the gauge variables is specified by recalling that the classical Yang-Mills
action involves the curl of $A_\mu$, $F_{\mu\nu}$.
Thus a lattice action ought to involve a product of
$U_\mu$'s around some closed contour. The trace of this product is gauge invariant,
so any lattice action which is a sum of powers of traces of $U$'s around closed loops
is a good gauge action. (Yes, this is arbitrary; wait a while $\dots$).

If we assume that the gauge fields are smooth, we can expand the link
variables in a power series in  $gaA_\mu's$. 
(This is called ``taking the naive continuum limit.'')
For almost any (planar) closed loop, the
leading term in the expansion will be proportional to $F_{\mu\nu}^2$.
We might want our action to have the same normalization as the continuum action.
This would provide one constraint among the lattice coupling constants.

The simplest  contour has a perimeter of four links. In $SU(N)$ we write our candidate action as
\bee
\beta S= \frac{2}{g^2}\sum_x \sum_{\mu>\nu}{\rm  Re \ Tr \ }( 1 - U_{P\mu\nu}(x) ).  \label{eq:Waction}
\ee
where
\bee
U_{P\mu\nu}(x)=U_\mu(x)U_\nu(x+\hat\mu a)
U^\dagger_\mu(x+\hat\nu a) U^\dagger_\nu(x)
\ee
This action is called the ``plaquette action'' or the
``Wilson action'' after its inventor.
(Lattice jargon: $\beta=2N/g^2$.)

Let us see how this action reduces to the standard continuum action.
Assume that the $U$'s are nearly the identity and expand
\bee
U_\mu(x) = 1 + ia A_\mu(x) - \frac{a^2}{2}A_\mu(x)^2+\dots .
\ee 
Further assume that the $A_\mu$'s are smooth, so
 \bee
A_\mu(x+\hat\nu a) = A_\mu(x) + a \partial_\nu A_\mu(x) + \ldots \label{2.8}
\ee
The plaquette becomes
\bee
U_P(x) = 1 + i a^2 F_{\mu\nu} +\dots
\ee
where as usual
\bee
F_{\mu\nu}= \partial_\mu A_\nu - \partial_\nu A_\mu + i[A_\mu,A_\nu] ,
\ee
so in the small $a$ limit the action becomes the usual continuum gauge action
\bee
\beta S = \frac{1}{4g^2} \sum_x \sum_{\mu>\nu} a^4 F_{\mu\nu}^2+O(a^2)  = \frac{1}{4g^2} \int d^4 x F_{\mu\nu}^2 
\label{2.9}
\ee
transforming the sum on sites back to an integral.

\begin{figure}
\begin{center}
\includegraphics[width=0.6\textwidth,clip]{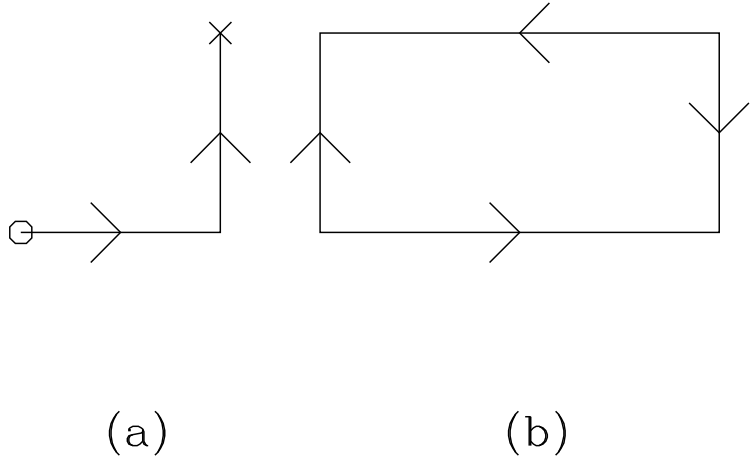}
\end{center}
\caption{ Gauge invariant observables are
either  (a) ordered chains (``strings'')
of links connecting quarks and antiquarks or
(b) closed loops of link variables. The arrows orient the product of link variables.}
\label{fig:figone}
\end{figure}

Note that if you want to do calculations directly with the lattice action, Eq.~\ref{eq:Waction},
integrating over the $U$'s,
 you don't need to gauge fix. The integration measure over any link variable $U_\mu(x)$ is finite.
The moment you want to do anything perturbative, you are working with the $A_\mu$'s and you
have to go back and pick up all the Fadeev-Popov
technology to deal with their flat directions.

\subsection{Confinement in strong coupling}
\begin{figure}
\begin{center}
\includegraphics[width=0.6\textwidth,clip]{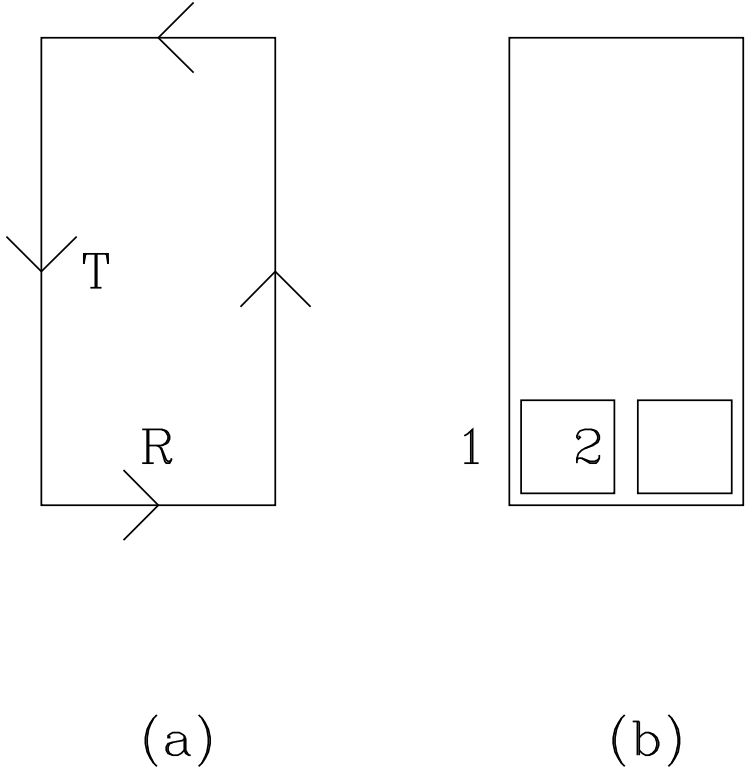}
\end{center}
\caption{An $R \times T$ Wilson loop (a)
and the pattern of tiling which occurs when it is evaluated in strong
coupling (b). }
\label{fig:figthree}
\end{figure}

In the same paper which introduced lattice gauge theory, Wilson 
showed that  in the strong coupling limit all gauge theories with the action of Eq.~\ref{eq:Waction}
exhibited confinement.  His order parameter was the Wilson loop.
It addresses the following question:
  In a world in which there are no light quarks, what is the
potential $V(R)$ between a heavy $q\bar{q}$ pair, separated by a distance $R$? 
 If the limit as $R$ goes
to infinity of $V(R)$ is infinite, we have confinement, if not, quarks are
not confined. $V(R)$ can be computed by considering the partition function in Euclidean 
space for gauge fields in
the presence of an external current distribution:
\bee
Z_J = \int [dU] \exp (-\beta S + i\int J_\mu A_\mu d^4 x) . \label{2.14}
\ee
 If $J_\mu$ represents a
point particle moving along a world line, it is a $\delta$-function on
that world line (parameterized by $\ell_\mu$):
\bee
i \int J_\mu A_\mu d^4 x = i \oint A_\mu dl_\mu . \label{2.15}
\ee
Gauge invariance implies current conservation and says that the
current line cannot end, and so the loop is a closed loop.  Make
it rectangular for simplicity, extending a distance R in some spatial
direction and a distance T in the Euclidean time direction.  This is the
famous Wilson loop. (See Fig.~\ref{fig:figthree}a.) $W(L,T)$ is the trace of a product
of $U$'s around the closed rectangular path.

$Z_J$ describes the loop immersed in a sea of gluons. We can think of
it as the partition function for a $q\bar{q}$ pair which is created at some time
$t=0$, pulled apart a distance R, and then allowed to annihilate at a time
$t=T$ later.

We want to find the difference in energy of the system  between when we include the pair and when it
is not present. We do this by measuring the response of the system as we stretch the loop a bit:
\bee
E_{q \bar q} = -\frac{\partial}{\partial T}
[ \ln Z_J - \ln Z_{J=0}] =   -\frac{\partial}{\partial T} \ln ( \frac{Z_J}{ Z_{J=0}}).
\ee
This is
\bee
E_{q \bar q} = -\frac{\partial}{\partial T} \ln{
\frac{\int [dU]e^{-\beta S}e^{i\int A \cdot dl} }{\int [dU]e^{-\beta S}}  }  
= -\frac{\partial}{\partial T} \ln\langle W \rangle_{J=0} 
 \label{eq:wloop}
\ee
where $\langle W \rangle $
 is the expectation value of the Wilson loop in the background
gluon field.  The behavior of $\langle W \rangle$ indicates confinement or not.

For example, suppose
$\langle W \rangle  \simeq \exp(-m(2R+2T)) $.
 This is called a ``perimeter law.'' It says $E=2m$.
The energy of the pair is independent of $R$, so the quarks can
separate arbitrarily far apart at a cost of finite energy.
 The quantity $m$ can be interpreted as the quark mass.
 However, if it happens that the Wilson loop shows  ``area law" behavior,
$\langle W \rangle  \rightarrow \exp(-\sigma RT) $, then $E=\sigma R$, and quarks are confined with a linear
 potential.  The parameter $\sigma$ is called the ``string tension.''
 In general, $\ln W$  \ will have some complicated form:
$ -{\rm ln}\langle W \rangle = \sigma RT +2m(R+T)
+ {\rm constant \ } + T/R + R/T \ldots $
 Nevertheless, if there is any area law term, it will dominate at large $R$ and $T$ and the
theory will exhibit confinement.
Notice that this confinement test fails in the presence of light dynamical
fermions. As the heavy $Q\bar Q$ pair are separated, at some point it will become
energetically favorable to pop a light $\bar q q$ pair out of the vacuum,
so that we are separating two color singlet mesons. The Wilson loop
will show a perimeter law behavior.

We can give an explicit demonstration of confinement in strong coupling for a U(1) gauge
theory \cite{Drouffe:1978dn}  (picking this gauge group just to keep the group theory simple).
 The Haar measure  for $U(1)$ is
\bee
\int dU = \int_{-\pi}^\pi \frac{d \theta}{2\pi} \label{2.17}
\ee
with  $\theta_\mu = ga A_\mu$,
and
\bee
Z= \int [dU] \exp (-\beta S)  \label{2.18}
\ee
where S is given by Eq.~\ref{eq:Waction}.
The plaquette is
\bee
U_{P\mu\nu}(x)=\cos(\theta_\mu(x) + \theta_\nu(x+\hat\mu a)
-\theta_\mu(x+\hat\nu a)- \theta_\nu(x)) .
\ee
  Now we write down two useful integrals:
\bee
\int_{-\pi}^\pi \frac{d \theta}{2\pi} e^{i\theta}  =0;\qquad
\int_{-\pi}^\pi \frac{d \theta}{2\pi} = 1  .  \label{eq:2.19}
\ee

Small $\beta$ is strong coupling and for small $\beta$ we may expand the exponential in
$Z$ as a power series in $\beta$:
\beea
Z &=& \int \prod_{links} \frac{d \theta}{2\pi}\prod_{plaquettes}(1 +
\sum_n \beta({\rm Tr}U_p(n) + {\rm h. \ c.}+ \ldots) \nonumber \\
&=& 1 + O(\beta^2) \nonumber \\
\label{2.20}
\eea
since all linear terms vanish. The ``dots'' are products of two, three, and more powers
of $\beta \sum_n U_P(n)$'s.

Now let's compute the expectation value of an L $\times$ T Wilson loop
(Fig. 2a):  There are LT plaquettes enclosed by it. The expectation value of the Wilson loop is
\bee
\langle W(L,T)\rangle =
\frac{1}{Z}\int \prod \frac{d \theta}{2\pi} \exp (i \sum_{WL} \theta)
\prod_{plaquettes}(1 +\sum_n \beta \cos  (\sum \theta) + \ldots)
\label{2.21}
\ee
It's easy to convince yourself that $W$ is proportional to
$\beta^{LT}$.  Pick one link on the boundary to integrate.
We need a term
in $\exp (-S)$ to contribute and cancel the phase of the boundary
link appearing in the Wilson loop (link 1 of Fig. 2b).  Now we
simultaneously need another
O($\beta$) term to cancel the phases of link 2 (of Fig. 2b).
 Only
if we ``tile" the whole loop as shown do we get a nonzero expectation
value. We are keeping the $N$th term in the sum, where $N=LT$. Each plaquette contributes a factor $\beta$ so
\bee
 \langle W(L,T)\rangle  \sim \beta^{LT}.
\ee
This means that
\bee
E_{q \bar q} = \sigma L 
\ee
where the string tension $\sigma= -\ln \beta$.
This is confinement by linear potential.

 The calculation can be generalized to all gauge groups. It means that,
in the strong coupling limit, any lattice gauge theory shows confinement.
This was a big result back in the day.
Wilson's calculation was the first
demonstration of confinement in a field theory in more than two
space-time dimensions.

\subsection{Hamiltonian formulation}
Unsurprisingly, there is a Hamiltonian for lattice field theories. It was invented by Kogut
 and Susskind \cite{Kogut:1974ag}
in 1974,
shortly after Wilson's paper. It can be derived from the path integral, but let's just cut to the chase and write 
it down.

As usual, to construct a Hamiltonian for a gauge theory it is necessary to fix gauge. The standard choice
is $A_0=0$ gauge. The Hamiltonian is
\bee
H = \frac{a^3}{g^2}\sum_x \{ \Tr \frac{1}{a^2}E^i(x)E^i(x) + \frac{2}{a^4} \sum_{i<j}\Re \Tr(1-P_{ij}(x) \}
\label{eq:ham}
\ee
where $E^i$ is the electric field in the $i$th spatial direction and $P_{ij}$ is the plaquette in the
$i,j$ plane.The dynamical variables are the links $U_i(x)$ and the electric fields. The Hamiltonian is
invariant under time-independent gauge transformations
\beea
U_i(x) &\rightarrow& V(x)U_i(x)V^\dagger(x+\hat i) \nonumber \\ 
E^i(x) &\rightarrow& V(x) E^i(x) V^\dagger(x) . \nonumber \\
\eea
And, there is a Gauss' law constraint,
\bee
C(x,t) = [D_i E^i] = 0 = \frac{1}{a^2}(E^i(x)-U_i^\dagger(x-\hat i) E^i(x-\hat i) U_i(x-\hat i))
\ee
which is time independent despite the notation.
You may encounter lattice gauge Hamiltonians in at least two places in the literature.
Oddly, both involve {\it classical} dynamics:

First, one of the standard simulation methods involves using the classical Hamiltonian to
generate new gauge configurations from old ones. The idea is to integrate Hamilton's
equations starting from Eq.~\ref{eq:ham}. If the paper you are reading mentions ``refreshed molecular dynamics'' or
``hybrid Monte Carlo,'' the authors' code is doing this. In this case, if your author is studying
a four dimensional Euclidean path integral, the spatial sums (the $i$ index in Eq.~\ref{eq:ham})
run from 1 to 4. The time in which the Hamiltonian is evolving is something external to the simulation.

The second place you might see this is if you are interested in real-time evolution of high-population 
(\ie classical) fields, for example in the early evolution of the quark gluon plasma.
Then $i=1$ to 3. There are a number of variations on this theme (for an early one, see Ref.~\cite{Krasnitz:1998ns}).
Some of them are not done over a regular grid; people who like metrics might want to read about them (on your own).

I will probably have more to say about the first case, later.

Back in the 70's, the literature on (quantum) Hamiltonian lattice QCD was probably as large as it was for
path integrals. This changed when Monte Carlo simulation came in. 
Working with the path integral, an
expectation value of some observable is
\bee
  \VEV{\cal O} = \left.\int [d\phi]\, {\cal O}(\phi) \exp[-S(\phi)]\right/
              \int [d\phi]\, \exp[-S(\phi)].
\label{eq:textbook}
\ee
For bosons, this is a classical integral. It can be studied using Monte Carlo methods,
which generate a sequence of $N$ random field
configurations $\phi^{(k)}$ chosen with a probability distribution given by 
\bee
  P(\phi) \propto \exp[-S(\phi)].
\label{eq:MCintro.equil}
\ee
The expectation value of the observable is just the simple average of
the observable over the ensemble of  configurations:
\bee
    \VEV{\cal O} = \frac{1}{N}\sum_{k=1}^N {\cal O}(\phi^{(k)}).
\label{eq:obsavg}
\ee
The uncertainty in the observation typically scales like $1/\sqrt{N}$. This means that
(at least in principle) the more data you collect, the better your answer. But working with Hamiltonians,
this is hard to do. Think about setting up a variational calculation, with some collection of
 $N$ basis functions. You get an answer. Does it depend on $N$?
You repeat the calculation with $2N$ basis functions.
Usually, you have to start over, to do this. The answer changes...
And we are not even talking about how much harder it is to do quantum mechanics than classical mechanics
(noncommutivity, for example).

Of course, there is a price to pay for working with a Euclidean functional integral -- no in or out states
means calculating transition amplitudes (or anything else involving an analytic continuation
back to Minkowski space) is difficult.

However, maybe quantum computing will let us do things which are hard with Euclidean path integrals,
like study real time evolution of a quantum system.
So it's worth knowing a few things about the quantum case.

The Hamiltonian can be rescaled to expose a strong coupling limit:
\bee
H = g^2\sum_x  \Tr \frac{1}{a^2}E^i(x)E^i(x) +  \sum_{i<j}\Re \Tr(1-P_{ij}(x)).
\label{eq:scham}
\ee
At large $g^2$ simply keep the electric term  and drop (or perturb in) the second term.
The quantum variables $E_i(x)$ and $U_j(x)$ inherit the commutation relations of $E$ and the vector potential $A$
and basically look like pairs of ladder operators.
For example, for a $U(1)$ group,  you can work in a $U-$ diagonal basis where
 $U(x)=\exp(i\theta(x))$ where $0<\theta<2\pi$,
 and then the electric field operator is $E= i\partial/\partial \theta$.
In the basis where $E$ is diagonal, its spectrum is discrete, a set of integers $m$ (think about the
 particle on the ring),
and $E^2 \sim m^2$. The Gauss constraint says that $m$ is conserved at all sites: we would say that
there is a string carrying $m$ quanta across the lattice, and that its flux is conserved by Gauss' law.
We could add external sources and sinks of flux (static charges) to terminate the string and 
then the energy of a pair of these
would scale with the distance between them -- linear confinement, with a constant string tension.
The $U$'s act like raising and lowering operators, so they move the $m$'s -- and so the
strings -- around. The $SU(2)$ case is worked out in
Kogut and Susskind: $SU(2)$ gauge theory is a collection of coupled rigid rotors.

Either way, path integral or Hamiltonian, we have a result: nearly all lattice gauge theories
exhibit confinement in the strong coupling limit. And it is kind of trivial: confinement is disorder
 (in path integral language)
or flux conservation (in Hamiltonian language).

This is an example of the use of a lattice calculation to
provide qualitative insight, which you couldn't get from a perturbative approach.
Of course, it isn't the last word!

\subsection{Getting real physics from a lattice calculation \dots}
This was a nice story, but even in 1974 it was a problematic one: it was a 
calculation in a quantum field
theory with a cutoff (the lattice spacing $a$) with a particular choice for a bare action (the Wilson action).
The gauge coupling $\beta$ is a bare coupling ($\beta(a)= 2N/g(a)^2$, 
re-inserting $a$'s everywhere to make the point). The formula
for the string tension, with the lattice spacing put back in, illustrates this: $\sigma= -(1/a^2)\ln \beta(a)$.
Dimensionally, $\sigma$ is a mass-squared, so $1/\sqrt{\sigma}$ is a length. We see that it
is proportional to the lattice spacing $a$, times a function of the bare coupling at scale $a$.

The lattice spacing is unphysical; we need to remove it from the calculation
and present cutoff-independent results.
Let's go back to basics. We imagine having some random bare theory involving a sum of bare operators $O_j$
and bare coupling constants $c_j(a)$. We keep all $O_j$ which are consistent with our symmetries. Everything
is written down in some convenient basis (for a gauge theory the $O_j$'s run over all possible closed loops).
We have a cutoff $a$, the lattice spacing. We scale everything with a dimension in units of $a$.
Our Lagrangian is
\bee
{\cal L} = \sum_j c_j O_j .
\label{eq:random}
\ee
Now we compute.
We measure a set of correlation lengths $\xi_k$ or masses $M_k$ in various channels.
For any random choice of $c_j(a)$, all the $\xi_k/a$ or $M_k a$ are order unity numbers.
(Even if there are special symmetries, this is true in a general case. You could have massless pions,
but the pseudoscalar decay constant would still be an order unity number times $1/a$.)

For us, the lattice spacing is an artifact. We want to  remove it from the calculation. 
I should say this sentence more carefully: we want to make predictions about physics on length scales
which are much larger than $a$, which are independent of $a$ and of the precise form of our
bare Lagrangian, Eq.~\ref{eq:random}.

The first thing we have to do is to get a theory which has physics on scales
which are much larger than $a$.
We do that by tuning the bare couplings,
to make the largest $\xi_k$'s diverge, $\xi_k/a \gg 1$. We observe empirically that
tuning some  $c_j$'s doesn't do much, while tuning other couplings does in fact drive the
$\xi_k$'s to large value.

We also observe that when the correlation lengths become large, physics becomes nearly
independent of what we chose to keep in  Eq.~\ref{eq:random}. ``Physics'' in this case means
dimensionless ratios of dimensionful quantities, since the dimensionful quantities that
come out of the actual calculations are still scaled by $a$.
Such a prediction could be a ratio of masses, and when one of the masses
became small, $a m_1 (a)\ll1$, we might see
\bee
[a m_1 (a)]/[a m_2 (a)] = m_1(0)/m_2(0) + {\cal O}(m_1a) + {\cal O}[(m_1 a)^2] +\dots.
\label{eq:SCALING}
\ee
The first term, the ratio $m_1(0)/m_2(0)$ is a real prediction;
 it is the $a\rightarrow 0$ or
(momentum) cutoff $\rightarrow$ infinity result. 
Everything else in Eq.~\ref{eq:SCALING} depends on the cutoff (on $a$).
(We call them ``lattice artifacts.'')

The last two paragraphs were presented with a kind of false naivete, because the situation
it describes is a bit more of what we expect rather than what we see. The end result,
Eq.~\ref{eq:SCALING}, is still the end result.

The theoretical background is the renormalization group, of course. The iconic presentation
for lattice students is still Wilson and Kogut \cite{Wilson:1973jj}.
In the language of the renormalization
group, the bare $c_j$'s are a set of coordinates in a more useful coupling constant space
of relevant, marginal and irrelevant operators, $\{ g_j \}$, and the correlation length diverges when
the relevant couplings are tuned to special values, ``to a critical point.''
Although I said ``correlation length,'' it is actually a whole set of correlation lengths
or inverse masses, which diverge,
all in some fixed ratio. We observe universal behavior, physics results (these ratios) independent
of most of the parameters in the bare theory.
If we are not quite at criticality, correlation lengths are finite but very large.
Ratios of correlation lengths will be nearly constant, with corrections which go like
$a/\xi$ to a power.
Remember, relevance, marginality, and irrelevance are defined with respect to the fixed point.
In renormalization group language again, we say, all our bare theories share a common set of
relevant operators (which we tuned to get where we are) but differ in their 
coupling to
irrelevant operators. Taking the continuum limit means not only $a\rightarrow 0$ but universality, independence
of predictions on precisely what went into the bare theory, Eq.~\ref{eq:random}.

So how do you actually make a prediction?
 You do a series of calculations
at ever smaller lattice spacings. Pick a set of bare parameters, compute (for me, that means simulate), measure.
How do you vary the lattice spacing? Change the bare parameters, repeat the calculation, 
look at what you got.  If Eq.~\ref{eq:SCALING} is what you see, then you are done.

Lattice people like to say
that one prediction of a mass determines the lattice spacing, when the value of that mass
 is fixed
by experiment. This is just the statement that $a= ma/m_{expt}$. One then uses this $a$ to make
predictions in energy units for other masses or dimensionful quantities, $m_2=m_2 a/a$. We 
say, ``at a lattice spacing of XX fm,
the ratio of the pion decay constant to the rho mass is YYY.''
This is of course an intermediate point on the way to a real prediction, which comes after
taking a limit, like in Eq.~\ref{eq:SCALING}.

\begin{figure}
\begin{center}
\includegraphics[width=0.7\textwidth,clip]{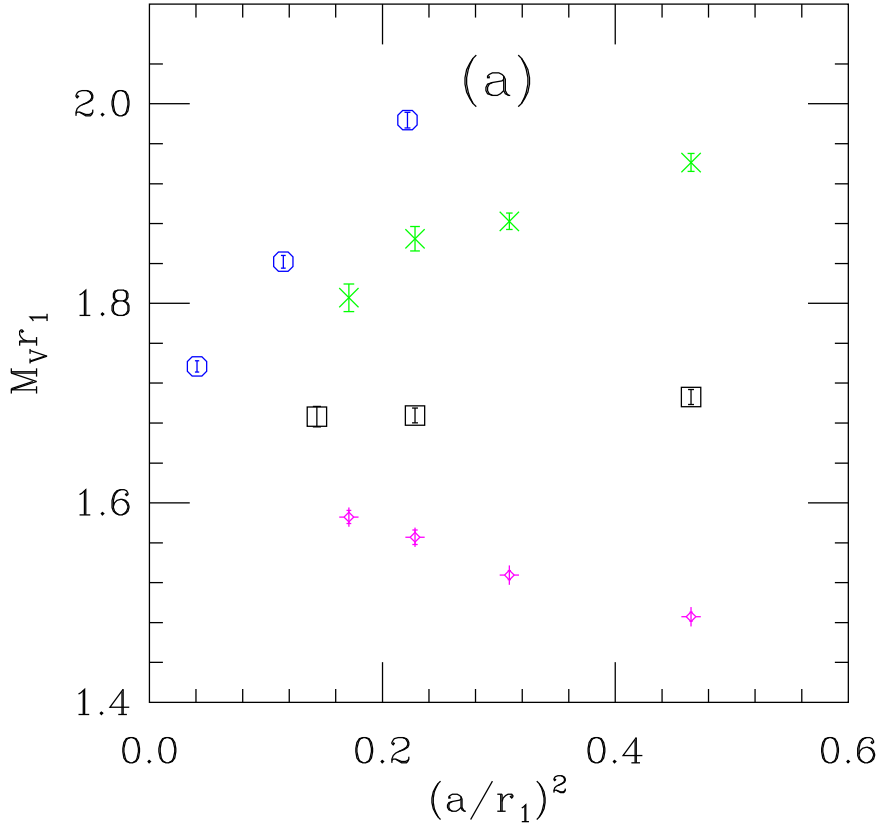}
\end{center}
\caption{An old picture (circa 2007)
illustrating Eq.~{\protect\ref{eq:SCALING}}.
The different symbols are calculations of the rho meson mass from different lattice discretizations,
in units of a length scale from the heavy quark potential, called $r_1$, which is about 0.3 fm.
You are supposed to imagine that all the symbols extrapolate to about the same value at zero lattice spacing.
\label{fig:scalingrho}}
\end{figure}

This is illustrated in an old (2007) picture I found, Fig.~\ref{fig:scalingrho}.
It shows predictions of the dimensionless ratio $m_V r_1$ versus lattice spacing from four
different sets of simulations with four different bare lattice actions. $m_V$ is the mass of the lightest
vector meson; $r_1$ is an inflection point in the heavy quark potential, which was used to
set the lattice spacing. Perhaps there was a common (universal?) limiting value at $a=0$.

Taking $a$ to zero by varying the bare parameters sounds like groping in the dark, 
and for any random theory, it would be.
But theories like QCD are special: they are asymptotically free.  Their 
 relevant couplings are the gauge coupling $g$ and fermion masses $m$. That's all.
 The system has a critical
surface in the space of all couplings
 that encloses a Gaussian fixed point at $g=0$ and $m=0$.  Tuning the two relevant couplings to zero
causes the correlation length, measured in units of $a$,
 to diverge. Since we are headed for a Gaussian fixed point, we know where to go.
Ever bigger  $\beta=2N/g^2$ is ever smaller $a$.
Relevancy, marginality, or irrelevancy are defined with respect to the Gaussian fixed point,
so it is easy to classify the operators in our action.

Another nice feature of  asymptotic freedom is that
 when the bare coupling is taken smaller
and smaller,
the short distance behavior of the theory becomes increasingly perturbative and 
hence increasingly controlled.
In particular, dimensions of fields and operators approach their engineering dimensions.
This allows us to parametrize the dependence of an observable on the cutoff scale.
  It is nearly given by naive
dimensional analysis. We can figure out which operators are irrelevant and which ones aren't.

 All lattice actions differ from
the expected continuum action of fermions coupled to gauge fields by the addition of
extra irrelevant operators.
We can see that for the Wilson gauge action itself.  The first terms in
the expansion of the plaquette in powers of the lattice spacing are
\bee
{\rm Tr}U_P = N + (1/2)a^4 O_4 + a^6\sum_i r_i O_{6i} +\dots
\ee
where the dimension-four term is
\bee
O_4 = \sum_{\mu\nu}{\rm Tr} F_{\mu\nu} F_{\mu\nu} 
\ee
and the three dimension-six terms are
\beea
O_{6a} &=& \sum_{\mu\nu}{\rm Tr} D_\mu F_{\mu\nu} D_\mu F_{\mu\nu} \nonumber \\
O_{6b} &=& \sum_{\mu\nu\rho}{\rm Tr} D_\mu F_{\nu\rho}  D_\mu F_{\nu\rho} \nonumber \\
O_{6c} &=& \sum_{\mu\nu\rho}{\rm Tr} D_\mu F_{\mu\rho} D_\nu F_{\nu\rho} .
\eea
Thus one would expect physical quantities
computed with the Wilson plaquette action to have ${\cal O}(a^2)$ lattice
artifacts. The dimension-six terms all break $O(4)$ invariance, but these are irrelevant operators,
so  these
symmetries  are expected to be restored in the continuum limit, as we work closer and closer
 to the Gaussian fixed point.

(The lattice has been around for a long time, so, especially in QCD simulations, there is
active work on designing actions whose artifacts are as small as possible.)

So a successful lattice calculation of an asymptotically free theory like QCD has a short distance part,
where the theorist (and his/her computer code) lives, with a small lattice spacing, a controlled field
content, and an action which is what you want to study (QCD) plus controllable dirt.
You (or rather, your computer) solves the system to give  predictions for long distance behavior.
You may have no idea what is going on
there, but because you live and work at short distances, you know what you are doing.

This is what Creutz's first computer simulations did, back in 1980 \cite{Creutz:1980zw}.
He did simulations varying the bare gauge coupling across its range.
He measured a nonvanishing string tension for all values of bare gauge coupling
and showed that the  weak coupling regime of lattice QCD was in the same phase as the strong coupling region.
We can take the weak coupling formula for the running coupling,
\bee
g^2(a) = \frac{1}{b \ln \frac{a_0}{a}}
\ee
and invert it,
\bee
a \sim a_0 \exp \left( -\frac{1}{bg^2(a)}\right).
\ee
This tells how a lattice mass $aM(g^2(a))$ should vary with bare coupling $g^2(a)$ in the weak coupling limit.
For Creutz, $M_H$ was the square root of the string tension, so he
 was able to see confinement (the string tension) occurring simultaneously with
asymptotic freedom.
Continuum QCD is confining.
Yes, this result is an empirical fact, nobody has ``proven'' that it occurs. So what? Nobody cares!

Another old issue: essentially all lattice regulated gauge theories confine in the strong coupling limit.
What about systems like QED (or pure $U(1)$ gauge theory)? One would presume that there would be some 
sort of critical
behavior someplace in the space of bare coupling constants, so that the strong and weak 
coupling phases are not analytically connected. People looked for these transitions in the early days
of lattice simulations, and usually found them. For example, four dimensional $U(1)$ lattice gauge theory
with the plaquette action has a transition (around $\beta=1$)
 between a strong coupling, electrically confining, 
magnetically screened
phase, and a weak coupling deconfined phase whose spectrum consists of a massless free photon. 

Just to end this lecture on a ``glass half empty'' note, what if you want to study a system 
which is not asymptotically free? Life is not so straightforward. Universality still (presumably) works:
if you can tune to a place where the correlation length diverges, you can still 
characterize the system in terms of relevant and marginal operators
with irrelevant corrections. But you have to search for the critical point.
And away from weak coupling, scaling dimensions of operators may be different
 from their engineering dimensions. It may not be possible to identify
relevant versus irrelevant operators in terms of their underlying field content. Worse, the system may happen to
lie in the basin of attraction of  fixed points other than the one you are looking for, or may be susceptible
to non-universal lattice-artifact phase transitions which depend on the particular choice of discretization.


\section{Where the bodies are buried -- all the parts of  typical lattice calculations}

In this lecture I want to give you an idea of how simulations are actually done. I'll start
with bosonic systems,
 then talk about fermions. I'll finish with general remarks: which systems are easy
 to simulate, which ones are hard or impossible.

\subsection{Simulating bosons}
Remember the goal: given a set of field variables $\phi_i$  defined on lattice sites or links $i$,
and an action function $S(\phi)$, we want to compute an expectation value
\bee
  \svev{\cal O}= \frac{1}{Z}\int [d\phi]\, {\cal O}(\phi) \exp[-S(\phi)] .
\label{eq:tb2}
\ee
The integral on the right hand side of this formula has a dimensionality proportional
to the number of lattice sites, so it is a lost cause to try to evaluate it exactly.
But the weight corresponding to most of the field values is vanishingly small,
so it is simply a waste of time to try to add them up. Instead, lattice people
use ``importance sampling'' (or ``Monte Carlo'' or ``Markov Chain Monte Carlo'') to approximate the answer.
These methods generate a sequence of $N$ random field
configurations $\{\phi^{(k)}\}$ with a probability distribution given by 
\bee
  P(\phi) \propto \exp[-S(\phi)].
\label{eq:equil}
\ee
The expectation value of the observable is then just the simple average of
the observable over the ensemble of  configurations:
\bee
    \VEV{\cal O} = \frac{1}{N}\sum_{k=1}^N {\cal O}(\phi^{(k)}) + O(\frac{1}{\sqrt{N}}).
\label{eq:MCMC}
\ee
How are the samples chosen? Typically, the $k$th configuration in Eq.~\ref{eq:MCMC} 
is generated from the $(k-1)$st one. (This is the ``Markov chain'' part of the name.) To
do this, you invent some
transformation rule $T(\{\phi\}  \rightarrow \{\phi' \}  )$. If it happens that
\bee
\exp(-S(\phi))T(\{\phi\}\rightarrow \{\phi'\} )= \exp(-S(\phi'))T(\{\phi'\}\rightarrow \{\phi\})   
\ee
(this is called ``detailed balance'') and if the $T$ algorithm is not too weird
 (it has to be ergodic, all possible
configurations have to be accessible in a finite number of steps from any starting one)
 then the probability distribution
generated by $T$ will eventually converge to $P(\phi)=\exp(-S(\phi))/Z$ and we 
obtain Eq.~\ref{eq:MCMC}.

One example of a $T$ function is the ``Metropolis'' algorithm \cite{Metropolis:1953am}.
Typically, it implemented running through the lattice variables in some sequence and
updating each one separately.  Call the field variable on site $i$ in the $k$th ensemble $\phi_i^{(k)}$.
For the Metropolis algorithm, to update a $\phi_i^{(k)}$,
compute the original action $S_o$. Make a change in $\phi_i^{(k)}$ to find a proposed value
 $\phi_i'$: if it were an element of
a set of discrete variables, pick a new discrete value. If it were a 
continuous variable, make some continuous transformation. For example, for an $SU(N)$ link $U$,
multiply by another $SU(N)$ matrix $V$; $U^{(k+1)} = V U^{(k)}$. Compute the new action $S_f$.
Now for the Metropolis rule: If $S_f \le S_o$, assign the new
$\phi_i^{(k+1)}$ to be  $\phi_i'$. If $S_f>S_o$,
make the replacement with probability $\exp(-(S_f-S_o))$. (That is, cast a random number $r$
uniformly distributed between 0 and 1, and make the change if $r<\exp(-(S_f-S_o)$)).
 If we can make the change, we say that we have 
an ``acceptance.'' If the proposed change is rejected, $\phi_i^{(k+1)}=\phi_i^{(k)}$.

A second kind of updating is called ``molecular dynamics.''
These algorithms basically exploit the
micro-canonical ensemble: they
use classical dynamics and the
ergodic hypothesis to obtain the desired statistical distribution. 
To keep this simple, imagine the path integral for a set of bosonsic variables $\phi$
with an action $S(\phi)$.
Introduce a fictitious momentum $p_n$
conjugate to $\phi_n$ at each lattice site $n$ and consider the
Hamiltonian
\bee
  H(p,\phi) = \sum_n \frac{p_n^2}{2} + S(\phi).
\ee
This Hamiltonian defines classical evolution in ``molecular dynamics
time'' $\tau$:
\bee
  \dot\phi_n  = p_n; \ \ \ \dot p_n = -\partial S/\partial \phi_n,
\label{eq:update}
\ee
where the dot denotes the $\tau$ derivative. 
(In practice, these equations are discretized with a time step $\Delta t$.) Starting from an initial
choice $[p(\tau = 0), \phi(\tau = 0)]$ these equations define a
trajectory $[p(\tau), \phi(\tau)]$ through phase space. The set of all
such trajectories is area-preserving. The corresponding classical
partition function is
\bee
   Z = \int [dp]\, [d\phi] \exp[-H(p,\phi)].
\ee
Integrating out the momenta returns us to our path integral, so  the quantum partition
function for $\phi$ can be evaluated using classical molecular dynamics.
According to the ergodic hypothesis, the probability of visiting a
point $\phi$ along the classical trajectory is proportional to
$\exp[-S(\phi)]$.  Expectation values of observables are then computed
by simply averaging over the molecular dynamics ``trajectory'':
\bee
  \VEV{\cal O} =
    \frac{1}{T}\int_{\tau_0}^{\tau_0 + T}d\tau\, {\cal O}[\phi(\tau)].
\ee
Variations of this procedure are called ``refreshed molecular dynamics'' (where the $p$'s are
periodically re-initialized as Gaussian random numbers) and ``Hybrid Monte Carlo,''
which has an Metropolis accept-reject step to correct the finite-time step integration of Eq.~\ref{eq:update}.

As a concrete example, consider the Ising model. The $\phi_i$ fields are spins, $s_i=\pm 1$
and
\bee
S = -\beta \sum_i \sum_\mu s_i s_{i+\mu} .
\ee
A simulation of the Ising model would go as follows:

Pick a value of $\beta$.
Begin by assigning all the spins some arbitrary value. The system could be ordered (all $s_i=1$)
or disordered (assign spins randomly), or anything
else you want, like a configuration you already generated at some different $\beta$.
 Now start running your update algorithm. Sweep
 through the lattice attempting to change all the spins. Do this for a while.
Monitor things that are easy to measure, like the average value of the spin, or the average internal energy, 
proportional to $1/V \sum_i \sum_\mu s_i s_{i+\mu}$. Your goal at this point is to 
bring the system into equilibrium, where Eq.~\ref{eq:equil} is true.

\begin{figure}
\begin{center}
\includegraphics[width=0.7\textwidth,clip]{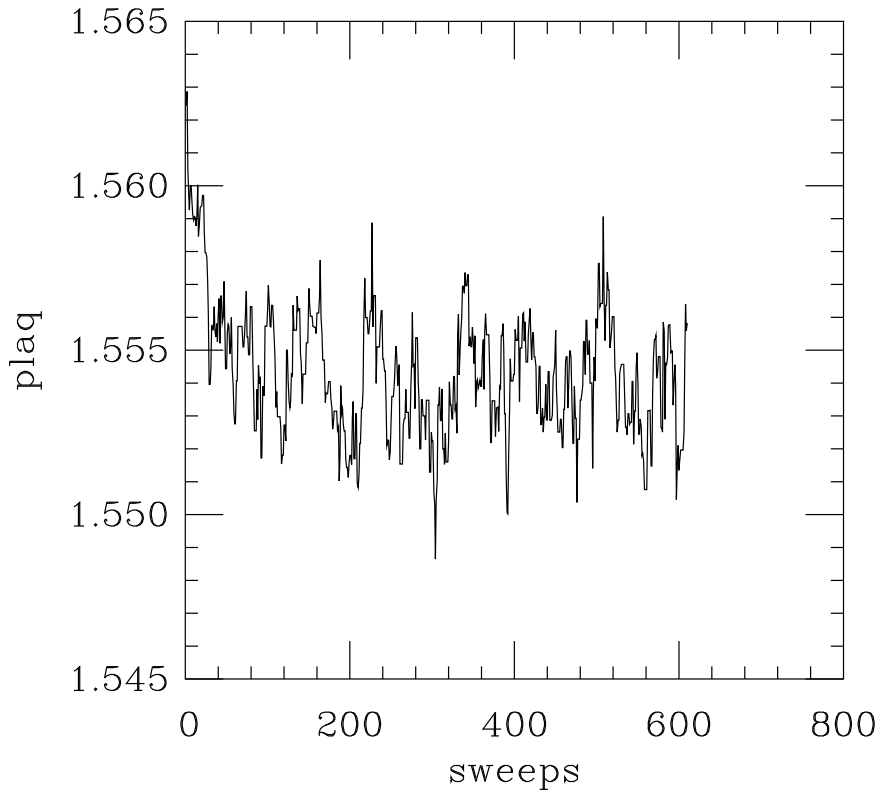}
\end{center}
\caption{Coming into equilibrium (in a QCD simulation): notice how the average plaquette falls before
appearing to fluctuate around a mean. Also notice the time autocorrelations.
\label{fig:running}}
\end{figure}

You can see this happening in Fig.~\ref{fig:running} (which is actually molecular dynamics for QCD).

When you think you are in equilibrium, you can begin to collect real data. However,
at this point you hit the biggest problem of Monte Carlo simulation: correlations.
They come in two kinds. 

The first kind is called ``time autocorrelations.''
In the Markov chain, new field values $\phi^{(k+J)}$ depend on old ones $\phi^{(k)}$.
This means that successive terms in the series $\phi^{(k)} \rightarrow\phi^{(k+1)} \rightarrow\phi^{(k+2)} \rightarrow
\dots\phi^{(k+N)}$
 are not independent. (Remember, you don't accept all the time.)
You can (probably) see these time autocorrelations by eye in Fig.~\ref{fig:running}.
A simple cure is to do measurements at long time intervals. This may not be feasible, however.
Life is finite!
 Monte Carlo
is basically diffusive, and the time correlation length in the data will typically scale with the
square of the spatial correlation length, $\tau \sim \xi^2$. This means that when the
correlation length becomes long, the simulation will freeze up. This isn't good; as you saw
last time, a long correlation length is where you want to be doing physics. (People
try to design updating algorithms to get around this.)

The second kind of correlation would exist even if there were no time autocorrelations.
Suppose you want to measure the mass of a hadron in QCD. You can do this by looking at some
correlation function with a source and a sink at two different points on the lattice.
The operators have the quantum numbers of the hadron. You expect the correlation function to fall exponentially
with distance, with the mass (inverse correlation length) characterizing the falloff: schematically,
\bee
C(t) = \sum_x\svev{O(x,t)O(0,0)} \sim \exp(-m_H t)
\label{eq:corrt}
\ee
You have a set of independent lattices. On each one of them, you measure
$\sum_x O(x,t)O(0,0)$ at many values of $t$. You then average them to recover
$C(t)$, which you then fit to $A\exp(-m_H t)$ to output $m_H$. But the individual $C(t)$'s at each $t$ value
came from the same underlying field configurations, so they are all correlated with each other.
People know how to deal with this (they do ``correlated chi-squared fits'') and now you know
to watch out for it.

So much for bosons. There are still a lot of things to go wrong, but let's move on.

\subsubsection{Fermionic Monte Carlo}
Lattice fermions are really annoying. First of all, it is tricky to write down a lattice
fermion action with the right number of degrees of freedom. (This is called the ``doubling problem.'')
Second, computers  can't deal with Grassmann variables and
so you have to integrate them out exactly before the simulation starts. This is hard to do
 if the fermion action is not a bilinear in the fermion degrees of freedom,
so let's restrict the discussion to that case. (If you are interested in four-fermion interactions,
you need to introduce auxiliary bosonic variables using some variation of the Hubbard-Stratonovich
transformation
\bee
\exp{-(\bar \psi \psi)^2 } =
\int d\phi \exp -{(\bar \psi \psi \phi + \phi^2)}
\ee
to get back to a quadratic form.)

Let's postpone the doubling issue to Sec.~\ref{sec:chiral}.

Computers can't do Grassmann algebra, so everybody in the lattice world first does the formal integration
over the fermions and then deals with the result.
 For single component fermions this gives a Pfaffian
of the fermionic action. For the kind of actions particle people deal with the multiple degrees
of freedom promote the result into a determinant.
Consider
full QCD with a single flavor of Dirac
fermion. If  its partition function is
\bee
   Z = \int [dU] [d\bar \psi ] [d \psi]\, \exp[-S_G(U)-\bar \psi D(U)\psi],
\ee
we integrate formally to make it
\bee
   Z = \int [dU]\, \exp[-S_G(U)] \det D(U) .
\label{eq:fermionmethod.fullQCD}
\ee
The determinant is nonlocal, so computing its change under a change in any single link of the
gauge field is very expensive.  The standard way to make the problem tractable has two parts. The first part 
is to simulate the determinant by introducing a scalar
``pseudofermion'' field $\Phi$, and making use of the formal identity
\bee
  \det D(U) = \int [d\Phi^* d\Phi]\exp\, [ -\Phi^* D^{-1}\Phi].
\label{eq:pf1}
\ee
This gives us an all-boson functional integral. But it also gets us into trouble.
The identity  requires all eigenvalues of the matrix $D$ to have a positive real part.
Unfortunately, the
eigenvalues of lattice Dirac operators are complex and their real
parts may not be positive-definite. Individual terms in the exponential can be
complex or carry a net negative sign. Then the exponential in
Eq.~(\ref{eq:pf1}) cannot be interpreted as a conventional probability
measure.

The solution of all fermion algorithms that I know of is
 a variation on the same idea:
invent a matrix whose determinant is the same as $\det D$, but whose eigenvalues
are real and positive-definite.
The way this problem is usually circumvented
 involves an explicit doubling with corrections to come
later.  For most fermion actions one can show that $\det D = \det D^\dagger$ 
(or $D^\dagger=\gamma_5 D \gamma_5$). Suppose you want to simulate two degenerate flavors.
The determinant  is $\det D^2 = \det D^\dagger D$, which has a real and (usually) positive spectrum.
just what we want. Our pseudofermion action has become
\beea
   Z 
     &=& \int [dU]\, \exp[-S_G(U)] \det[ D^\dagger(U) D(U)]  \nonumber \\
      &=& \int [dU d\Phi^* d\Phi]\, \exp[-S_G(U) - \Phi^*(D^\dagger D)^{-1}\Phi].
\label{eq:fermionmethod.phiaction}
\eea
This is fine for QCD if we think that two degenerate flavors
(up and down quarks) are a good approximation to Nature. Going away from two flavors, people work with
fractional powers of $\det D^\dagger D$,
\bee
(\det D)^{n_f} = (\det D^\dagger D)^{n_f/2}
\ee
and then introduce more complicated pseudofermion actions to handle the fractional power.
These are variations of rational approximations with formulas like
\bee
\frac{1}{x^p} = \frac{\prod(x+a_n)}{\prod(x+c_m)} = \sum_n \frac{b_n}{x+c_n}
\ee
This is how the four tastes of staggered fermions are reduced to a single flavor.
(We are getting too technical, so let's move on.)

Regardless of what we do, the pseudofermion algorithm is still nonlocal. To compute the change in
the action if  a single
bosonic variable (a gauge link) is changed requires O(volume) operations.
So we can't use something like Metropolis. Instead, all fermion Monte Carlo I know use
molecular dynamics algorithms. 
Eq.~\ref{eq:update} tells us why these algorithms are used:
we can change all the link variables at once, hence we only have to recompute $D^{-1}$ (or $(D^\dagger D)^{-1}$ once.
So the algorithm costs order(volume), not order (volume${}^2$).
The price is, the time step $\Delta t$ has to be kept small or the time integration goes unstable.


\subsection{It's time to measure something}
Now pretend you have a system you want to simulate: QCD, the Ising model, ${\cal N}=4$ SYM, $\dots$.
You have written a code. What physics can you do?

The short answer: you can measure any observable $O$ which can be written as
\bee
  \svev{\cal O}= \frac{1}{Z}\int [d\phi]\, {\cal O}(\phi) \exp[-S(\phi)].
\ee
(You might already be in trouble -- think about trying to find the free energy, or the entropy.
But let's go on.) There are two generic kinds of observables, global averages and correlation functions.

If you have a new system, the first things you measure are global averages. 
You  want to map  the phase structure of your model in terms of its bare parameters.
 You will be particularly interested in second (or higher) order phase transitions.
These are places where the correlation length diverges; they are where
continuum behavior exists.

Think about an Ising model.
At small $\beta$, the average spin $\svev{s_i}$ will average to zero
and at large $\beta$ the system will be ordered. Something might happen, in between these values.
There might be an abrupt change in the qualitative behavior of the system at some $\beta_c$.
Is this a phase transition? If so, is it a first order transition, or does it look smooth?
(First order boundaries support separated phases, sometimes.)
Usually, it is easy to tell that something is going on, but it can be hard to say more.

And of course there is a catch: usually a finite volume system does not support a true phase transition,
or true symmetry breaking. You will probably have to do simulations on several volumes to
try to sort out what is going on. (More about this in Sec.~\ref{sec:ising}.)

Global quantities are usually not the most interesting ones. Masses and matrix elements
are computed from unamputated correlators like $\svev{J(x_1)J(x_2)}$ or $\svev{J(x_1)J(x_2)J(x_3)}$.
Let's look at some examples, starting with the mass of a particle.

\begin{figure}
\includegraphics[width=0.6\textwidth,clip]{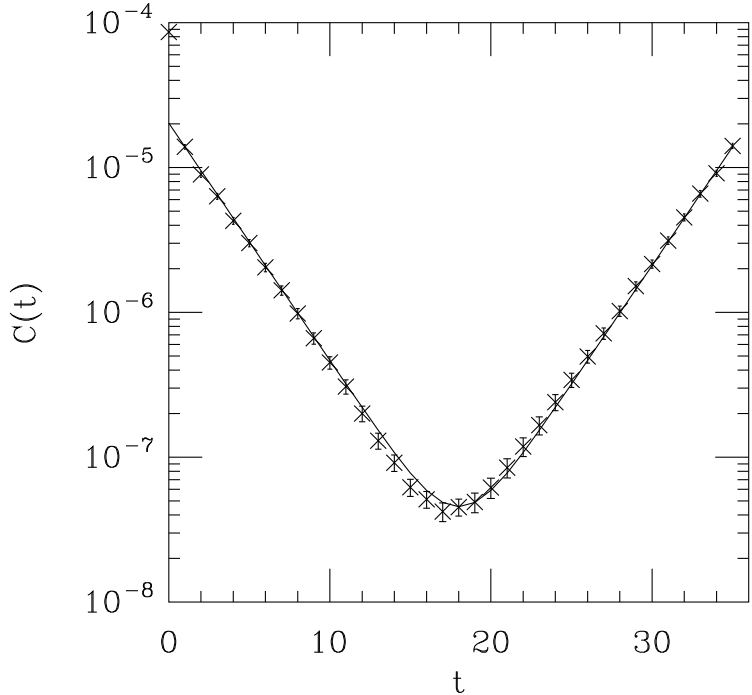}
\caption{
An atypically nice lattice correlator and its
fit. In this case $N_t a = 32 a$. The discussion in the text was a bit too casual about boundary effects,
which turn the exponentials into hyperbolic cosines. }
\label{fig:prop}
\end{figure}

Consider the expectation value of Euclidean correlator
\bee
C_{ij}(t) =\svev{ 0| O_i(t) O_j(0) |0 }
\label{CT}
\ee
where the
 operators create some state of interest from the
vacuum.  Making the replacement
\bee
O_i(t)=e^{Ht}O_ie^{-Ht}
\ee
and inserting a complete set of energy eigenstates into Eq.~(\ref{CT})
yields
\bee
 C_{ij}(t) = \sum_{n} \bra{0} O_i \ket{n}  \bra{n} O_j \ket{0}
   e^{-E_nt}.
\ee
The correlator is a sum of exponentials. 
(We have
assumed that the spectrum of energy eigenstates is discrete.)
If the operators had vacuum expectation values, the
leading term in the sum would be a constant. If that is not the case, 
the correlator is a sum of falling exponentials. The lightest state ($n=1$)
in the channel contributes
the smallest exponential and will dominate the correlator at large $t$.
  If the operator projects onto  zero three-momentum,
$E_1$ is the mass of the lightest state it excites.
Fig.~\ref{fig:prop} gives an example of such a correlator,  a pion ($u \bar d$) propagator.

This already tells you:
\begin{itemize}
\item Masses of states which do not have vacuum quantum numbers (pions or 
protons in QCD but not the scalar glueball).
are easier to measure 
\item The lightest state in a channel is easier to study than an excited state.
\item A clever choice of $O_i$ can enhance your signal, a bad one can kill it.
\end{itemize}

\begin{figure}
\begin{center}
\includegraphics[width=0.6\textwidth,clip]{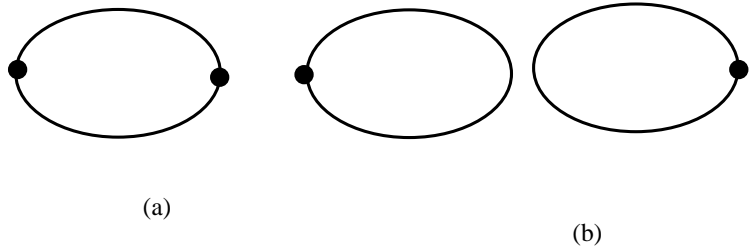}
\end{center}
\caption{ Connected (a) and disconnected (b) quark diagrams
corresponding to the two terms in Eq.~(\protect{\ref{eq:generalcorr}}).}
\label{fig:conndisc}
\end{figure}

Most of the interesting observables in theories like QCD involve valence fermions.  The
interpolating operators create or annihilate fermion fields from the
vacuum.  Let's suppose we wanted to measure the mass of a meson.
 Then we might consider measuring a correlation function
\bee
C(t) = \sum_x \langle J(x,t) J(0,0) \rangle
\ee
where $J(x,t) = \bar \psi(x,t) \Gamma \psi(x,t)$
and $\Gamma$ is a Dirac matrix.   The intermediate states
$|n \rangle$ that saturate $C(x,t)$ are the hadrons that the current
$J$ can create from the vacuum: the pion, for a pseudoscalar current,
the rho, for a vector current, and so on.  Written in terms of the
fermion fields, the correlator is
\bee
C(t) = \sum_x \langle 0 | \bar \psi_i^\alpha(x,t)  \Gamma_{ij}
\psi_j^\alpha(x,t)\bar\psi_k^\beta(0,0)
\Gamma_{kl}\psi_l^\beta(0,0) | 0 \rangle
\ee
with a Roman index for spin and a Greek index for color.  We
contract creation and annihilation operators into quark
propagators,
\bee
\langle 0 | \psi_j^\alpha(x,t) \bar \psi_k^\beta(0,0) | 0 \rangle
 = G_{jk}^{\alpha \beta}(x,t;0,0).
\ee
There are two $\psi$ fields and two $\bar \psi$ fields in the meson
correlator, so there are two ways to pair them in the contraction.
One way pairs the $\psi$ in
the source with the $\bar \psi$ in the sink and vice versa.
The other way pairs  $\psi$ with the $\bar \psi$ in the source, forcing
the same contraction in the sink.   See
Fig.~\ref{fig:conndisc}. Also remembering sign changes from
interchanging Grassmann variables, we  have
\bee
C(t)
 = \sum_x {\rm Tr}\left[ G(x,t;0,0) \Gamma G(0,0;x,t) \Gamma \right]
- \sum_x \Tr \left[ G(x,t;x,t) \Gamma\right]
    \Tr \left[G(0,0;0,0) \Gamma \right]
\label{eq:generalcorr}
\ee
where the trace runs over spin and color indices.  If the meson is a
flavor nonsinglet, the second contraction gives zero.  Baryon
correlators are constructed similarly. 
 We see that the space-time
``Feynman rule'' for Fig.~\ref{fig:conndisc} associates a valence
quark line from $(x^\prime,t^\prime)$ to $(x,t)$ with $G_{jk}^{\alpha
\beta}(x,t;x^\prime,t^\prime)$, but does not display gluon and sea
quark lines.

The computation of the first (quark-line connected) term in
Eq.~\ref{eq:generalcorr} typically proceeds as follows.  The propagator
equation for the Dirac operator
$(D+m)G(x,t;x_0,t_0)=S(x_0,t_0)$
 is solved with a source  at
$(x_0,t_0)$ to get $G(x,t;x_0,t_0)$ for all sink locations $(x,t)$.  This
calculation is done for each source color and spin. 
This is a sparse matrix inversion problem; the solution is usually some iterative
technique (see Numerical Recipes for more). Then the propagators are sewn together.

\begin{figure}
\begin{center}
\includegraphics[angle=90,width=0.7\textwidth,clip]{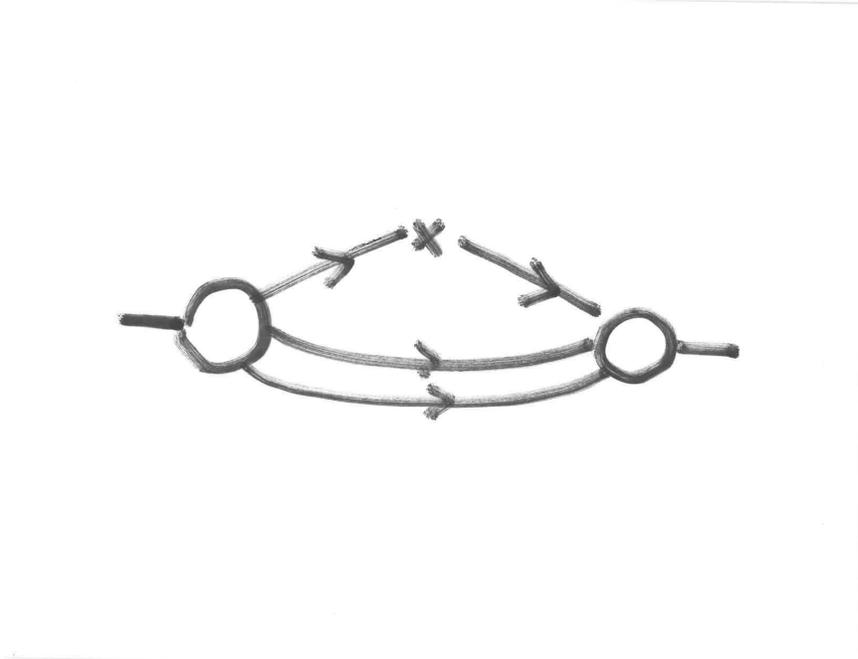}
\end{center}
\caption{An unamputated diagram used in computing a form factor.
\label{fig:ff}}
\end{figure}

There is a similar story for matrix elements. Recall that the pseudoscalar decay constant
$f_P$ comes from
\bee
\langle 0 | J^5_\mu |P(k)\rangle = -ik_\mu \sqrt{2} f_P e^{-ikx}.
\label{eq:fP}
\ee
This matrix element can be computed from the two-point correlator
\bee
C_{JO}(t)= \sum_x \langle 0 | J(x,t) O(0,0) | 0 \rangle 
\rightarrow  e^{-m_P t} \frac {\langle 0|O|P\rangle\langle P|J|0\rangle}{{2m_P}}
\label{eq:CURR2}
\ee
A second calculation of
\bee
C_{OO}(t)= \sum_x \langle 0 | O(x,t) O(0,0) | 0 \rangle
\rightarrow  e^{-m_P t} \frac {{|\langle 0|O|P\rangle|^2}}{{2m_P}}
\label{eq:cj1}
\ee
is needed to extract $\langle 0|J|P\rangle$, which is accomplished by
fitting the two correlators with three parameters, $m_P$, $\langle 0|O|P\rangle$, and $\langle 0|J|P\rangle$.
 Quantities like $\svev{h|O|h'}$ involve diagrams like
Fig.~\ref{fig:ff}.

One final issue with matrix elements: typically, they are
scale and scheme dependent. Phenomenologists usually want an $\overline{MS}$
number at some fiducial scale. It is necessary to convert the
result in the lattice scheme to the continuum one.
 Sometimes this is done using perturbation theory;
 often, another lattice calculation is performed,
to determine the matching factor nonperturbatively.

\subsection{Back to the big picture, again}

You have your favorite theory, can I simulate it for you? 
I think this question does not have a crisp answer,
but here goes:

Let's start out with ``impossible systems.'' These are ones where $\exp(-S)$ is not real
 and positive, so it cannot be used as a probability for importance sampling.
 ``Impossible''
is not such a nice word, so people say that these systems have a ``sign problem'' or a ``phase problem.''
Examples of such systems include QCD at nonzero chemical potential, QCD with a theta term, $N=4$ SYM, and condensed matter systems
at  nonzero chemical potential or away from half-filled bands.

Actually, there are a lot of people working on impossible systems. Most of the literature is about 
successfully simulating them,
rather than actually doing physics with them. My impression is that many interesting systems have
sign problems, but the issues are all different.

Many  systems have real actions. Then the issue is, 
how much continuum physics can you get out of your simulation?

Easy systems to simulate include spin models and pure gauge theories, both bosonic systems.
I would say that these days QCD or QCD-like systems with favorable fermion content
 (like two degenerate not-too-light flavors)
are ``easy.''  What  makes
QCD hard are things like taking the fermion masses to their physical values, 
taking the volume to infinity, taking
the lattice spacing to zero. The issue with the fermion mass is that fermion algorithms 
involve repeatedly 
computing fermion
propagators, that is,
 inverting $D+m$. This is typically done with some iterative sparse matrix inversion procedure, 
and the problem is
that the cost of these algorithms scales like the conditioning number of the matrix, which 
is the ratio of the largest
to smallest eigenvalues. This scales inversely with the fermion mass.
Critical slowing down pushes the cost of a QCD simulation toward scaling as a larger inverse power of the
quark mass and slightly more than the volume. (Compare the old discussion in Ref.\cite{Jansen:2003nt}.)
But these days, people (mostly)  know what they are getting themselves into, before they start.

Many systems are reasonably straightforward to simulate but hard to analyze. Near conformal systems
are a good example. The ones which have been most studied are asymptotically free but with slowly running couplings.
They came out of the technicolor game. Briefly, people wanted to find systems which were like QCD in
 that they confined, but
had slowly running couplings and large scaling dimensions, for working phenomenology. For about ten years after 2008
lattice people studied $SU(N)$ gauge theories with
many fermionic degrees of freedom. It was easy to see that the theories ran slowly. (This was done by measuring
various lattice observables which could be interpreted as a scale dependent coupling constant.) The problem
was telling whether the coupling always ran slowly, or if the analog of the beta function really had a zero.
The main issue is that it's not possible to simulate over a wide range of length scales in a single simulation.
If a system has a slowly running coupling, then if it is weakly interacting at short distance, it is
weakly interacting at long distance. If it is strongly interacting at long distance it is strongly 
interacting at short distance. And if the system is strongly interacting at short distance, 
you don't know what you are doing.
The situation as of a few years ago is described in my  review \cite{DeGrand:2015zxa}.

There is only a small lattice literature about SUSY. 
I know about
simulations of ${\cal N}=1$ and ${\cal N}=4$ supersymmetric Yang - Mills theory
 in space - time dimension $D=4$, and various models lower dimensions. 
Motivation is an issue: what are crisp questions
we could answer? Maybe we could test AdS/CFT ideas by direct simulation in strong coupling ${\cal N}=4$
super Yang-Mills.

Of course, one has somehow to evade the problem that supersymmetry is an extension of the
usual Poincar\'e algebra and so it is broken completely by naive discretization.
However, my understanding is that this is a mostly solved problem, in principle. But I could be wrong.
All I know about this subject is from the
review article Ref.~\cite{Catterall:2009it}.

${\cal N}=1$ is a  system of adjoint Majorana fermions
coupled to gauge fields. 
 The supersymmetric limit is the limit of vanishing fermion mass. This is not impossible, just hard.
Some representative papers include \cite{Fleming:2000fa,Giedt:2008xm,Endres:2009yp,Kim:2011fw,Bergner:2013nwa}.
These are confining systems so the interesting thing to look for is the SUSY-related degeneracy in the spectrum --
the lightest states should be a degenerate multiplet of a scalar (a mixture of meson and glueball),
a pseudoscalar, and a spin - 1/2 fermion (a gluino-gluon bound state). A recent paper,
Ref.~\cite{Ali:2019agk}, shows this.

${\cal N}=4$ is much trickier, because of  the scalars. Any naive discretization of scalars will introduce
a hierarchy problem: the scalars will get a mass which is inversely proportional to the lattice spacing. 
An intricate
construction described in \cite{Catterall:2012yq}
 allows one to simulate a theory with a single scalar supercharge. The other
fifteen supercharges of  ${\cal N}=4$ are broken by the lattice discretization.
It is believed that the situation is like the loss of rotational invariance in a more conventional lattice
system: the breaking of the symmetry is due to irrelevant operators. This means that  these
supersymmetries are recovered in the continuum limit. Exactly how to do that in an efficient way
 is presently a research problem. (And ${\cal N}=4$ has a phase problem, so maybe it is impossible, after all.)
I worked on this for a while: see \cite{Catterall:2012yq,Catterall:2014vka} for what we did.
 David Schaich's recent review \cite{Schaich:2018mmv} is the best recent summary I know.

People have had better luck with lower dimensional SUSY systems.
A partial list of these studies includes
 \cite{Honda:2013nfa, Hanada:2013rga,Honda:2011qk,Ishiki:2009sg,Ishii:2008ib,Ishiki:2008te}.

\section{Chirality on the lattice\label{sec:chiral}}

Lattice QCD people spend a fair amount of time thinking about chiral symmetry.
 Spontaneous chiral symmetry breaking explains why the pions are light;
explicit chiral symmetry breaking (through the quark masses) explains why the pions
 are not massless, and why the kaons are heavier than the pions. The presence of the anomaly for the
 flavor singlet axial current tells us that the eta and eta-prime are heavier still.
Knowing the quark mass dependence of operators (which comes from chiral symmetry) helps
us take simulation data at unphysical quark masses and make predictions at the physical values.

QCD is a vector gauge theory, the two chiralities of fermions couple equally to the gluons.
The Standard Model is a chiral gauge theory: left handed fermions and right hand fermions couple differently.
So the second motivation to think about lattice chiral symmetry is that it
would be nice to have a nonperturbative regulator for the Standard Model. 

The simplest lattice fermions have issues with chiral symmetry. The choices we have are to
work with fermion actions which are chiral but doubled (naive or staggered fermions)
or undoubled but with explicit order $a$ violations of chiral symmetry (Wilson or clover fermions).
  Issues have consequences. 
Wilson fermions have to be fine tuned; the bare quark mass is additively renormalized,
  $m_q= Z m_q^{lattice} + C$,
so when you start a simulation you don't really know where you are. Operator mixing is
a more serious issue. If lattice symmetries are different from continuum ones, then
desired matrix elements can be contaminated by mixing with operators of different chiral structure.
For staggered fermions, loss of full chiral symmetry means the loss of degeneracy in would-be
chiral multiplets. The pions are not all degenerate.
In both of these formulations the anomaly picks up order $a$ or $a^2$ lattice artifacts.

To be honest, these are issues that QCD people know about, and people have learned to live with them.
And yet, it would be nice to do better.

There are  lattice actions with exact chirality.
Around the time I was writing these lectures, I got interested in reading the
topological insulator literature (two reviews are Refs.~\cite{Shankar,Tong:2016kpv}. 
Big parts of it smelled very familiar, like
things my friends did 20-25 years ago.
So I wrote David Kaplan. He agreed,  there are connections between
the kind of lattice actions that we particle physicists use, and topics in the topological insulator game.
But there are many parts that are not shared between the two communities.
There are some things that we lattice people know, that I have not seen in the
articles I have read. There are probably other things that we do not know, and they know.
There is a project for someone (maybe one of you) to do, to synthesize what the two fields have done.
I can't do it. The language is too different.
I am not on top of all the connections. But I can tell you about lattice chirality, in the language
we use in lattice QCD.

\subsection{The doubling problem}

Let's start with the issue: simple ideas don't work.
A continuum free massless fermion with a Dirac operator $D=\gamma_\mu \partial_\mu$ has a propagator
\bee
S(p) = \frac{-i \gamma_\mu p_\mu}{p^2}.
\ee
It has a single pole, at $p_\mu=(0,0,0,0)$, and it is chiral in the sense that $[\gamma_5, D]_+=0$
and that one could project out various helicity states from the propagator in the standard way.

It is hard to do any of that on the lattice. In fact, there is a  famous ``no-go'' theorem about doubling
and chirality, due to
Nielsen and Ninomiya \cite{Nielsen:1980rz,Nielsen:1981xu}. In detail, the theorem assumes
\begin{itemize}
\item
A quadratic fermion action $\bar \psi(x)i H(x-y)\psi(y)$, where $H(p)$ is
continuous and well behaved. It should behave as $\gamma_\mu p_\mu$ for
small $p_\mu$.
\item
A local conserved charge $Q$ defined as $Q=\sum_x j_0(x)$, which is quantized
(\ie $Q$ doesn't change across the Brillouin zone)
\end{itemize}
The statement of the theorem is that, once these conditions hold,
$H(p)$ has an equal number of left handed and right handed fermions
for each eigenvalue of $Q$: this is doubling.
 The upshot is that we will have to find clever ways of proceeding
if we want  chiral symmetry breaking to be a result of dynamics, not how
we discretized the fermions.

Conventional lattice fermions are either chiral but doubled (naive or staggered fermions)
or undoubled but with order $a$ or $a^2$ violations of chiral symmetry (Wilson or clover fermions).

Naive lattice are chiral but doubled. Their action is
constructed by replacing the derivatives by symmetric
differences. It is
\bee
   S_L^{\rm naive} =  \bar \psi D^{\rm naive} \psi =
  \sum_{n,\mu} \bar \psi_n \gamma_\mu  \Delta_\mu\psi_n
 + m \sum_n \bar \psi_n \psi_n , \label{2.24}
\ee
where the lattice derivative is
\bee
\Delta_\mu \psi_n = \frac{1 }{ {2a}}
(\psi_{n+\hat\mu} - \psi_{n-\hat\mu}).
\ee
The free propagator is easy to construct:
\bee
\frac{1}{ a} S(p) = (i \gamma_\mu \sin p_\mu  a + ma)^{-1}
 = \frac{-i \gamma_\mu \sin p_\mu a + ma}{\sum_\mu \sin^2 p_\mu a + m^2 a^2} ,
\label{2.25}
\ee
 The massless propagator has poles at $p=(0,0,0,0)$,
$ap=(\pi,0,0,0),$ $(0,\pi,0,0)$, \dots, $(\pi,\pi,\pi,\pi)$,  in all the
corners of the Brillouin zone.  Thus our action is a model for sixteen
light fermions, not one. 

The 16 naive fermions can be shown to decouple into four groups of four ``tastes,'' and it is possible
to simulate only a single set of four tastes. This is called a ``staggered fermion.''
Staggered fermions maintain some chiral symmetry, but at the cost of introducing doublers. A
single staggered fermion corresponds to four degenerate flavors in the naive continuum limit.
Staggered fermions have a single component per site, so a full Dirac spinor is spread around on the lattice.

One way to avoid doubling would be to alter the dispersion relation so
that it has only one low energy solution.  The other solutions are
forced to $E \sim 1/a$ and become very heavy as $a$ is taken to
zero.  The simplest version of this solution, called a Wilson fermion,
adds an irrelevant operator, a second-derivative-like term
\bee
  S^W   =
   -\frac{r}{ {2a}}\sum_{n,\mu}\bar \psi_n(\psi_{n+\hat\mu} -2 \psi_n
  +\psi_{n-\hat\mu} ) \simeq - \frac{ar}{2} \bar \psi D^2 \psi \label{2.26}
\ee
to $S^{\rm naive}$.  The
propagator will become
\bee
\frac{1}{ a} S(p) = \frac{-i \gamma_\mu \sin p_\mu a +
   m a -r \sum_\mu (\cos p_\mu a -1)}
 {\sum_\mu \sin^2 p_\mu a + [m  a-r \sum_\mu(\cos p_\mu a -1)]^2} .
\label{2.27}
\ee
It remains large at $p_\mu \simeq (0,0,0,0)$, but the ``doubler
modes'' are lifted at any fixed nonzero $r$ to masses that are order
$1/a$, so $S(p)$ has one four-component minimum. Unfortunately,
the Wilson term is not chiral.

This discussion makes lattice fermions sound like a disaster. Reality is not so extreme --
a better word than ``disaster'' is ``annoyance.''
Modern QCD simulations have a lot of parts: small lattice spacing, big volume, complicated operators.
People rarely try to make one part of the simulation perfect; it's better to do many things reasonably well.
 The actions most people use in simulations
are highly improved versions of staggered or Wilson fermions, tuned to reduce lattice artifacts
while remaining computationally efficient.

\subsection{Chirality from five dimensions}

The first, and still most used, path to a chiral fermion is through the fifth dimension.

Domain wall fermions are our version of edge states in topological insulators.
They are a variation on the old Jackiw-Rebbi \cite{Jackiw:1975fn} story, that a massless fermion
can sit on the side of a soliton (at the place where a scalar field $\phi(s)$ interpolates between
two asymptotic values). We want  four dimensional chiral symmetry, so
the $s$ dimension is a fifth dimension, and the side
of the soliton is our four dimensional world. 

Our classic papers are by
Kaplan \cite{Kaplan:1992bt} and Shamir \cite{Shamir:1993zy}.
The lattice version of the Callan-Harvey paper \cite{Callan:1984sa} is
described in Ref.~\cite{Golterman:1992ub}
and  Ref.~\cite{Kaplan:1995pe} is also worth a look.

It is simple: here is the story in continuum variables. We imagine a free Dirac operator
\bee
D_5 = D_4 + \gamma_5 \partial_5 - M(s)
\label{eq:scharge}
\ee
in a five-dimensional Euclidean world, labeling  the usual coordinates $x_\mu$ with $\mu=1$ to 4,
 and a fifth dimension labeled by $s$. The operators are
$D_4 = \gamma_\mu \partial_\mu$ and $\partial_5 = \partial/\partial s$.
 The ``mass parameter'' $M(s)$ is assumed to vary with $s$,
interpolating between $-M$ and $M$.
 $M$ is supposed to be very large, $M\sim 1/a$ when we go back to the lattice.
We look for Euclidean space solutions of the Dirac equation $D_5 \chi = 0$, writing
$\chi(x,s) = \exp(ipx)u(s)$. Momenta are $p=(iE, \vec p)$ in Euclidean space,
where $p^2 = -E^2 + \vec p^2 = -m^2$. If $D_5 \chi = 0$, then
\bee
[\gamma_5 \partial_5 -M(s)]u = -i \not\! p u.
\ee
Squaring the equation,  we find
\bee
Hu= [-\partial_5^2 + V(s)]u = m^2 u.
\label{eq:susy}
\ee
where $V(s)= M^2 + \gamma_5 \partial_5 M$. The solutions to this equation include ones with nonzero
$m^2$, paired in chirality, plus a chiral zero mode localized around the $s$ where $M(s)=0$.

This is how lattice people tell their story, but students at this Tasi will recognize that it is
basically a SUSY quantum mechanics story, with the ingredients relabeled. Let's do the relabeling.
Our left and right handed states are the analogs of boson and fermion 
states $\ket{b}$ and $\ket{f}$.
The derivative of the superpotential is $W'(s)=M(s)$, $\gamma_5$ plays the role of $\sigma_z$.
Eq.~\ref{eq:scharge} defines a supercharge (actually multiplied by $\sigma_1$)
\bee
Q= \sigma_1 p - \sigma_2 W'=\sigma_1(p-i\sigma_3 W')
\ee
where $p=-i\partial_5$. The Hamiltonian $H$ in Eq.~\ref{eq:susy} is $Q^2$. States with
nonzero $m$ are paired; $\ket{b}$ and $\ket{f}=(Q/m)\ket{b}$ are degenerate. The Witten index
$(-1)^F$ is just the difference in the numbers of zero modes of the two chiralities.
They have wave functions
\bee
 \left( \begin{array}{c} \phi_R(s) \\ \phi_L(s) \end{array} \right) =
 \left( \begin{array}{c} \exp(W(s)) \\ \exp(-W(s)) \end{array} \right).
\label{eq:witt}
\ee
The usual story in the lattice literature is that $W'(s)$ is a function which interpolates
between minus and plus infinity as $s$ goes to plus or minus infinity, so that only
one of these modes is normalizable. The survivor is a chiral mode sitting near 
a zero of $W'(s)$.

Part of the lattice literature stops here. We have a set of chiral fermion zero modes
and chirally paired nonzero modes. We need to decide which of these five dimensional
fermion modes correspond to the four dimensional ones we wanted to study, but this
is just a technical complication.
 Deep in the infrared,
only the chiral modes contribute to physics and the massive, paired modes are just
 physics at the cutoff scale.

SUSY people would say that this
 is a general situation, independent of any specific details about
the superpotential. 
The topological insulator people would say that the zero modes are topological modes.

The word ``topology'' is not emphasized in the old lattice
literature,
 but lattice people knew: all this story was stable against changes in $W(s)$.
A complication that you may not have thought about is that if you want to code up
one of these actions, the variable $s$ cannot run over an infinite range. It's easiest to
think about $s$ as a ring of circumference $L$. Then, $W(s)$ is periodic and 
the Witten index has to be zero; there are modes of each chirality localized at different 
places around the ring. The engineering goal for a QCD simulation
 is to hang onto two zero modes and
combine them into a single massless Dirac spinor; what happens most of the time is that
the Witten index stays zero and the two would-be zero modes get lifted to some
(hopefully) small value. Lattice people call this value a ``residual mass.''

This takes us to the domain  wall fermion of the lattice literature. It was
introduced by Shamir\cite{Shamir:1993zy}.  Rather than a kink in $M(s)$
 (that is Kaplan's story),
it has  a five-dimensional
space $0<s<L$, with Dirichlet boundary conditions at both ends. The superpotential is
simply $W(s)=Ms$ where $M$ is a constant. Then the BPS (zero energy)  modes  sit at the ends
of the space,
\bee
 \left( \begin{array}{c} \phi_R(s) \\ \phi_L(s) \end{array} \right) =
 \left( \begin{array}{c} \exp(-M(L-s) \\ \exp(-Ms) \end{array} \right).
\label{eq:shamir}
\ee
Going back to the lattice, $s$ is discretized, and there are 
$L_5$ sites in the fifth dimension. 
The Dirac part of $D_4$ is some undoubled lattice action like the Wilson fermion action,
to exclude doubling from the start.
The gauge fields
go into $D_4$, so they are  just replicated on each four dimensional slice of the five
 dimensional lattice. Mass terms couple left hand fermions to
 right hand ones, so
to package the two chiral modes into a single Dirac
spinor, Shamir added terms $m( \bar\Psi(x,0) P_- \Psi(x,L_5-1) + \bar \Psi(x,L_5-1) P_+ \Psi(x,0))$ to
the Dirac operator. After repeating the mode
expansion,  the sum of positive and negative chirality
edge modes can be replaced by an action for a single Dirac particle
\bee
S = \int d^4 x\, [\bar \psi(x)(D_4 + \mu)\psi(x)]
\ee
where the Dirac mass is $\mu$ is proportional to $m$.

A few more technical remarks. (1) If you are interested in Chern-Simons currents, 
check out Ref.~\cite{Golterman:1992ub}.
(2) In the lattice setup, the
Wilson term part of $D_4$ gets lumped with the $M$ parameter of
the five dimensional action. This prevents the exact separation of
variables between the fifth dimension and the other four.
  For any finite $L_5$ domain wall actions are not exactly chiral,
although they are much more chiral than a generic Wilson action.
People tune their actions to reduce $L_5$ while keeping chiral violations small.

And a final remark in case lattice people are reading these lectures:
At this Tasi there were several lecture series about anomalies. The theme
of nearly every series was that the anomaly could be tamed by considering
a system in one higher dimension,  where the real physical variables lived
on a boundary of the higher dimensional system. How the system is extended into the 
extra dimensions was arbitrary. As a lattice person myself, watching these lectures,
I could not help wondering: we have domain wall and overlap fermions. There are highly optimized
codes for simulating them. The algorithms we use were typically constructed in some
way that was strongly motivated by lattice considerations. Could we do things differently?
Is there something for us in the formal lectures about anomalies at this year's Tasi?

\subsection{The Ginsparg - Wilson relation -- lattice chirality in four dimensions}
Another way to evade the Nielsen - Ninomiya relation is to change the rules.
The Ginsparg - Wilson relation \cite{Ginsparg:1981bj}
 replaces the continuum definition of chiral symmetry,
$[\gamma_5, D]_+ =0$, by
\bee
 0= \gamma_5 D + D \gamma_5 - \frac{a}{r_0}D \gamma_5 D.
\label{eq:GW}
\ee
$a$ is the lattice spacing; $r_0$ is a constant. One could equivalently replace 
the usual chiral rotation $\delta \psi = i \epsilon \gamma_5 \psi$, $\delta \bar \psi = i \epsilon \bar
\psi \gamma_5$ by either
\bee
\delta \psi = i\epsilon \gamma_5\left(1-\frac{a}{2r_0}D\right) \psi; \qquad
\delta \bar \psi = i\epsilon  \bar \psi \left(1-\frac{a}{2r_0}D\right)\gamma_5
\ee
or
\bee
\delta \psi = i\epsilon \gamma_5\left(1-\frac{a}{r_0}D\right) \psi; \qquad
\delta \bar \psi = i\epsilon\bar\psi\gamma_5.
\label{eq:ovrot}
\ee

 Dirac operators which obey the Ginsparg Wilson relation
know about the index theorem; they have chiral zero modes (plus paired nonzero modes of opposite
chirality). They know about the anomaly: for smooth gauge fields,
\bee
\svev{\partial_\mu J^{\mu 5}} = -N_f \svev{Q_T(x)} = -N_f\svev{ \frac{g^2}{16 \pi^2}
    \epsilon^{\alpha\beta\mu\nu}F^a_{\alpha\beta}(x)F^a_{\mu\nu}(x)} .
\label{eq:anomaly2}
\ee
Violations of continuum chiral Ward identities, from the last term in Eq.~\ref{eq:GW},
are just contact terms. This means that they are not important in practical calculations
(say in relations among $N-$point functions).

Blocking only hides symmetries, it does not remove them.
Ginsparg and Wilson came up with their relation by performing a real space renormalization group transformation
on a continuum chiral fermion action.
 This was in 1981. The paper was ignored until 1997 (11 citations),
when it was rediscovered  by Peter Hasenfratz while he was cleaning out his desk.
Now it is renowned (over 1000 cites). The
reason the paper was lost  was that Ginsparg and Wilson
did not have an explicit formula for a Dirac operator which obeyed Eq.~\ref{eq:GW},
only an implicit RG formula.
 This was provided by
Narayanan and Neuberger, who came at chiral symmetry studying a system with an infinite number of regulator fields.
 Their action is called the ``overlap action,''
\cite{Neuberger:1997fp}:
\bee
 D= \frac{r_0}{a}\left[1 + \frac{d-r_0/a}{\sqrt{(d-r_0/a)^\dagger (d-r_0/a)}}\right]
\label{eq:overlap}
\ee
or
\bee
 D= \frac{r_0}{a}\left[1 + \gamma_5 \epsilon(h)\right]
\label{eq:matstep}
\ee
Here $d$ can be any undoubled lattice fermion action. In Eq.~\ref{eq:matstep}, $h=\gamma_5(d-r_0/a)$ and
$\epsilon(h)$ is the ``matrix step function.''

I have used this action in simulations, It is complicated to code and expensive to evaluate, but for somebody
working alone,
it nice to use since there are a lot of lattice artifacts you don't have to check.

I have not found any analogs of overlap fermions in the topological insulator literature.
 Look at Eq.~\ref{eq:ovrot}:
the physics is that a chiral rotation is not performed on a single site; it smears the 
fermion over some $O(a)$ range.

Amazingly, domain wall fermions know about the Ginsparg - Wilson relation.
It is the effective action for fermions confined to the edges of the fifth dimension.
One way to show this is to consider the propagator between sites on the surface
as was done by L\"uscher  \cite{Luscher:2000hn}. Another way is to integrate out
the fermions in the bulk. Details for this can be found in Ref.~\cite{Brower:2012vk}.
The fermion determinant is
the product of a determinant of an approximate Ginsparg-Wilson fermion and a
determinant for the massive bulk modes. The approximation becomes exact as the number of sites in the
fifth dimension becomes large.
The action for the nearly chiral modes is
\bee
 D= \frac{r_0}{a}\left[1 + \gamma_5 \tilde \epsilon(h)\right]
\label{eq:matstepD}
\ee
where $\tilde \epsilon$ is an approximation to the step function, like
\bee
\tilde \epsilon(h) = \frac{(1+h)^L - (1-h)^L}{(1+h)^L + (1-h)^L}.
\ee

And this is a good place to mention topology. For QCD people, this is almost exclusively
a code word for
quantities related to  $F \tilde F$ -- the topological susceptibility, the eta prime, axion physics.
There is no simple lattice expression in terms of gauge fields (link variables)
 for an $F \tilde F$ whose integral is quantized;
the best hope is to invent quantities whose integrals approach an integer in the limit
of smooth gauge fields. This  is done with some smeared out lattice approximation to $F \tilde F$.
(Another alternative is to define an index by counting zero modes of an overlap operator.)
Simulating QCD at fixed winding number is possible; in fact, people worry when the
topological charge does not tunnel frequently during a simulation.
Simulating QCD-like systems at nonzero theta angle is difficult; there is a sign problem.

Finally, chiral gauge theories. This is a big mess. With a non-chiral lattice fermion, the lattice
gives you vector fermions. The doublers couple to gauge fields just like the $p=0$ fermions, but
they flip their chirality ($p_\mu$ goes with $\pi/a-p_\mu$).
You can't get rid of these ``mirror'' fermions without breaking gauge symmetry.
Refs.~\cite{Golterman:2000hr,Golterman:2004qv} describe approaches along these lines.
 The old Eichten - Preskill \cite{Eichten:1985ft} idea was to make the ``mirror'' heavy by giving the mirrors
strong interactions. They would form composites and get massive. One lattice study looked at this \cite{Golterman:1992yha}
-- it didn't work.

 L\"uscher
\cite{Luscher:2000hn} gives a fairly complete overview of the subject. He
 described   \cite{Luscher:2000zd} the construction of a
 $U(1)$ chiral gauge theory, but  (as far as I know) nobody has ever simulated it.

From a domain wall perspective, there is a one chirality on the right side of the fifth dimension
and the other chirality on the other side. Could one make  chiral fermion on one side invisible to the gauge fields?
This was the recent idea of Kaplan and Grabowska \cite{Grabowska:2015qpk,Grabowska:2016bis}.
 It also seems to have faded out.

I am absolutely not an expert in this topic! What I do know is, if you think you can construct a 
lattice chiral gauge theory, you are going to have
to code it up  and do simulations, if you want to convince people that your idea works.


\section{Case studies}

\subsection{The three-dimensional Ising model \label{sec:ising}}
\begin{figure}
\begin{center}
\includegraphics[width=0.8\textwidth,clip]{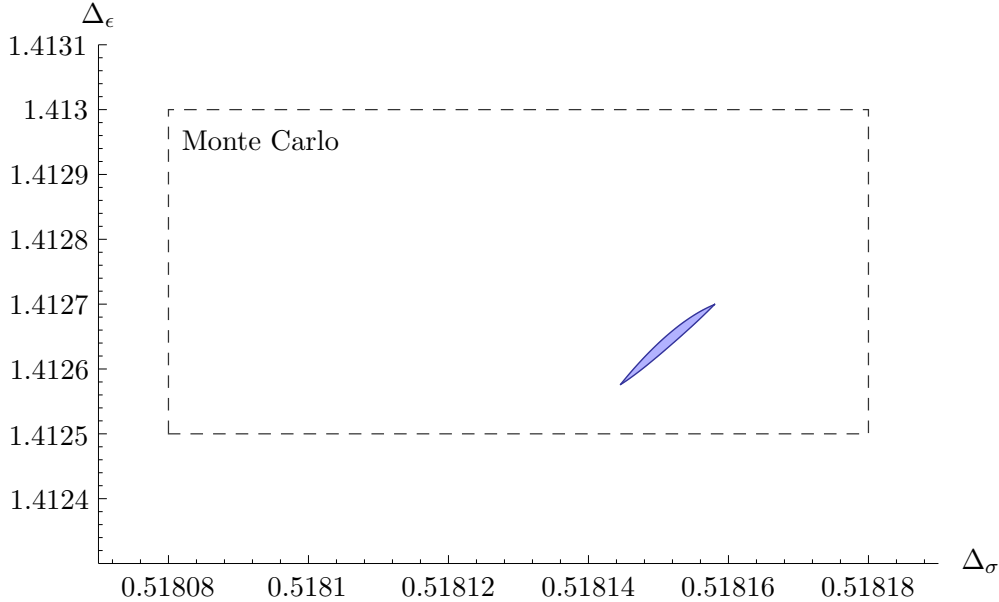}
\end{center}
\caption{Bound on $(\Delta_\sigma,\Delta_\epsilon)$ in a unitary 3d CFT with a $Z_2$ symmetry 
and two relevant scalars $\sigma,\epsilon$ with $Z_2$ charges $-,+$ from
Ref.~ {\protect{\cite{Simmons-Duffin:2016gjk}}}.
 The allowed region is the blue sliver. 
The dashed rectangle shows the $68\%$ confidence region from Monte Carlo determinations of 
Ref.~{\protect{\cite{Hasenbusch:2011yya}}}}.
\label{fig:dds} 
\end{figure}

Figure 30 of David Simmons-Duffin's Tasi 2015 lectures  \cite{Simmons-Duffin:2016gjk}
on the conformal bootstrap has
a plot of the two leading exponents of the three-dimensional Ising model, with a comparison
of conformal bootstrap results with Monte Carlo. I reproduce the figure in Fig.~\ref{fig:dds}.
This doesn't look like a positive picture for me to discuss, until you look up the citations
for the actual work: David's is Ref.~\cite{Simmons-Duffin:2015qma} from 2015, and the Monte Carlo is
five years earlier, by Hasenbusch, Ref.~\cite{Hasenbusch:2011yya}, on the arXiv in 2010. So how
did Hasenbusch do it?

There are a number of ways to get exponents out of a Monte Carlo simulation. Hasenbusch used 
a technique called ``finite size scaling.'' The idea is old and in textbooks
 (see Ref.~\cite{Cardy:1996xt}), but so is almost
everything else I have written about, and anyway, maybe you don't know it.

The physics motivation is that singularities in thermodynamic
quantities only happen in systems when the lengths $L$ of all the
dimensions to go to infinity. In the thermodynamic limit, fluctuations are correlated over a size
roughly equal to the correlation length $\xi$, and fall off as $\exp(-r/\xi)$. As long as
$\xi/L$ is small we have exponentially damped corrections of order $\exp(-L/\xi)$ due to edge effects.
(We use these to measure some quantities in some QCD simulations.)

But if $\xi \approx L$ we have problems. Let's imagine thinking about
a slab in $d=3$, thinner in height $H$ than in extent $L$. When $\xi \ll H$ the system thinks
it is 3-dimensional. For $\xi \gg H$ it will think it is two dimensional. This is called
``crossover behavior.''
In real experiments this is hard to see, but simulations are cleaner and $L$
(or $H$) becomes another parameter to tune.

Generally, in a simulation, one takes $L$ fixed in all dimensions. When $\xi \approx L$ the system
 is ``zero dimensional'' and there are no
singularities. Let's consider what happens to a susceptibility, which in an infinite system is 
expected to scale like
\bee
\chi \sim \xi^{\gamma/\nu}
\ee
(statistical mechanics conventions, the correlation length scales as $\xi \sim t^{-\nu}$ for $t=|T-T_c|$).
In a finite system we expect to see
\bee
\chi = \xi^{\gamma/\nu}\phi(\xi/L)
\ee
where $\phi(u)$ is a scaling function with $\phi(u)$ going to a constant as $u\rightarrow 0$.
On dimensional grounds we can trade $\xi$ for $L$:
\bee
\chi= L^{\gamma/\nu} \hat \phi(L/\xi)
\ee
or, with  $\xi \sim t^{-\nu}$, $L/\xi=Lt^\nu$, so, rewriting the scaling function,
\bee
\chi=L^{\gamma/\nu} \tilde \phi(L^{1/\nu}t).
\label{eq:fss}
\ee
If no dimensions are infinite, $\tilde \phi$ must be an analytic function of $t$. 
Thus $\tilde \phi(v)$ is smooth; it will have a peak at some $v=v_0$ of width $v_1$.
 Thus if we plot $\chi$ versus $T$ we expect to see
\begin{enumerate}
\item A peak at $L^{1/\nu}(T-T_c)=v_0$ (or $T=T_{peak}$, so $T_{peak}=T_c + v_0 L^{-1/\nu}$)
\item A width of $\Delta T = v_1 L^{-1/\nu}$
\item A height scaling like $L^{\gamma/\nu}$
\end{enumerate}
Item (1) implies that the apparent $T_c$ is shifted; items (2) and (3) that the peak in $\chi$ rises and narrows with $L$.

\begin{figure}
\begin{center}
\includegraphics[width=0.6\textwidth,clip]{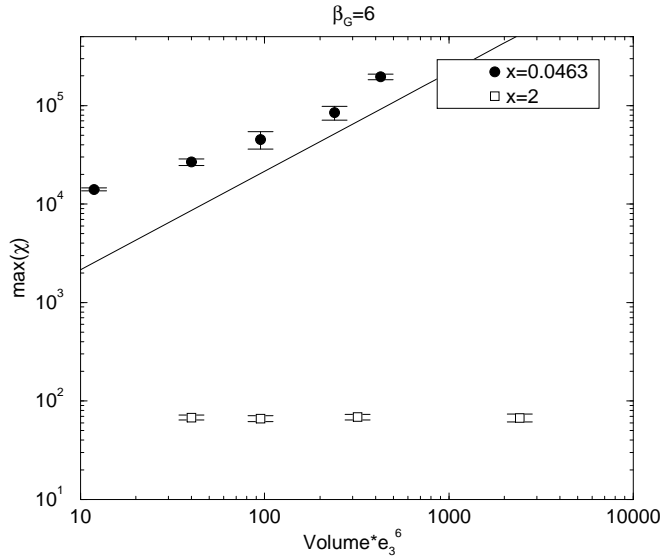}
\end{center}
\caption{The maximum of the  susceptibility $\chi$ in the $3-d$ Abelian Higgs model, 
as a function of volume,
from Ref.~{\protect \cite{Kajantie:1997vc}}. The straight line is $\sim V$. }
\label{fig:suscmax}
\end{figure}

There is an example of this, mentioned  in a paper
about duality transformations in three dimensional topological insulators
 by Metlitski and Vishwanath \cite{Metlitski:2015eka}.
They refer to a lattice study of the three-dimensional Ginzburg - Landau model
($3-d$ Abelian Higgs model) by Kajantie, Karjalainen, Laine, and Peisa \cite{Kajantie:1997vc}.
The issue is that the system has a line of phase transitions; what is the order of the transition?
Fig.~\ref{fig:suscmax} shows a plot of the maximum of the matter field susceptibility at two
bare parameter values. The physics is that if the peak in the susceptibility is measured
at the location of what would be a second order transition in infinite volume,
it would scale with volume like Eq.~\ref{eq:fss}. If
the peak sits on a first order boundary, it scales as the volume $V$ \cite{Fisher:1982xt}. It looks as if the
black points identify a first order transition, and the white points identify something else.

I've used this in my own work, for the correlation length itself:
\bee
\xi_L = L f(L^{1/\nu} t)   ,
\label{eq:fss2}
\ee
and Anna Hasenfratz and friends have done this with a nonleading term, \cite{Cheng:2013xha}
\bee
  \label{eq:expansion}
  \xi_L = L F_H(x)\left\{1 + g_0 t^{\omega} G_H(x) + O\left(g_0^2 t^{2\omega}\right)\right\}.
\ee
The first term is the leading  expression, Eq.~\ref{eq:fss2}, $x=L^{1/\nu} t$,
 and the long expression in brackets accounts for the leading corrections to scaling.

\begin{figure}
\begin{center}
\includegraphics[width=\columnwidth,clip]{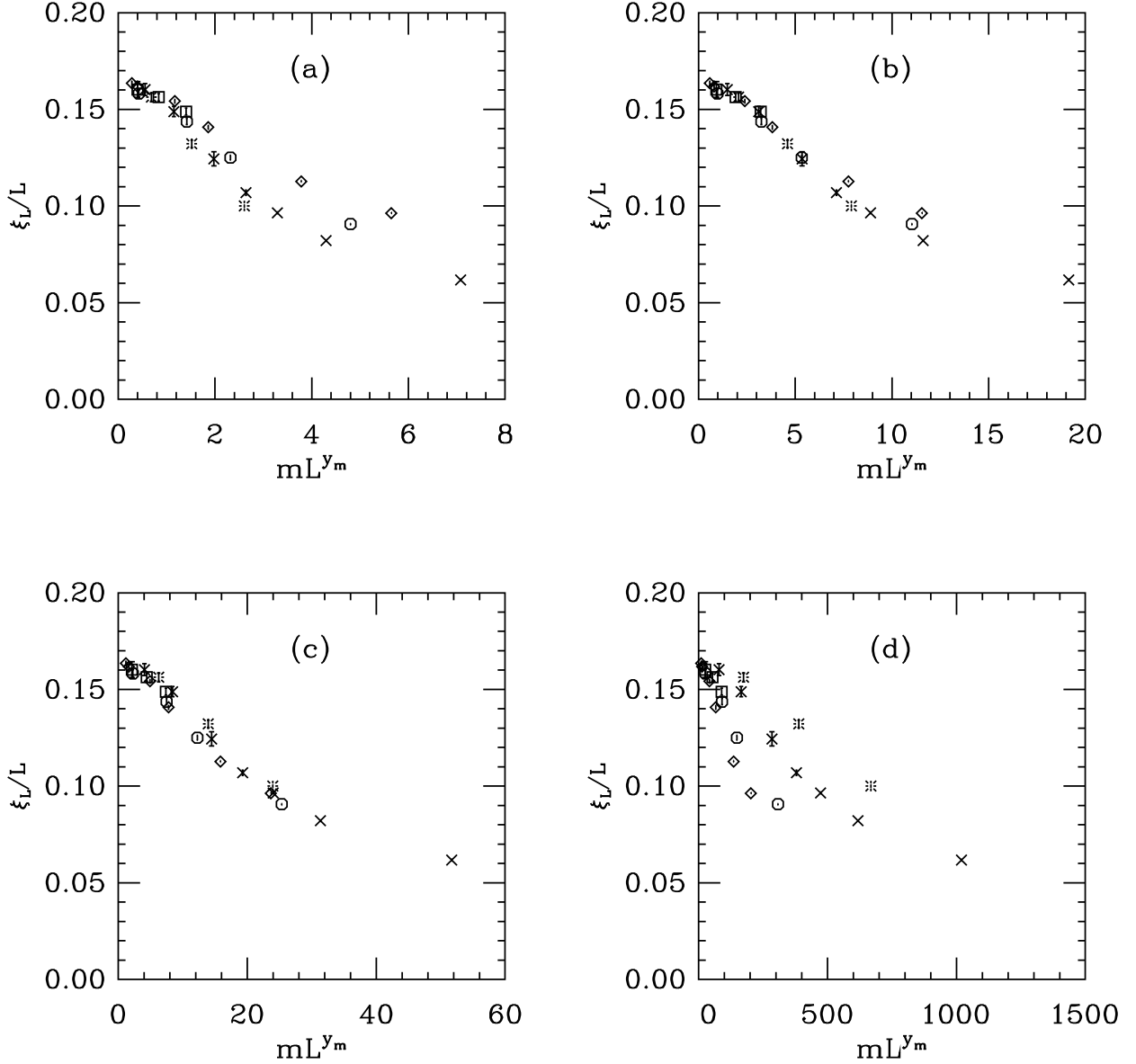}
\end{center}
\caption{
Curve collapse in $SU(3)$ gauge theory with $N_f=2$ symmetric-representation fermions, from
{\protect{\cite{DeGrand:2009hu}}}.
Plots of $\xi_L/L$ vs $m_q L^{y_m}$ at $\beta=5.2$ for four choices of $y_m$:
(a) $y_m=1.0$, (b) $y_m=1.4$, (c) $y_m=1.8$ (d) $y_m=3.0$.
Plotting symbols are for different simulation volumes,
diamonds, $12^3\times 6$ ($L=6$);
octagons, $12^3\times 8$ ($L=8$);
squares, $16^3\times 8$ ($L=8$);
crosses, $12^4$ ($L=12$);
bursts, $16^4$ ($L=16$). Curve collapse seems to be best in panels (b) and (c).
\label{fig:alltest}}
\end{figure}

So a way to find exponents in a Monte Carlo simulation is to measure some quantity, like a susceptibility,
for many $L$'s and for many $t$'s per $L$, and then try to do curve collapse by varying the exponent.
Anna and I were doing this for near-conformal systems. We plotted
 $\xi_L/L$ vs $L^{y_m} m_q$ for many $L$'s, and varied $y_m$. Under this variation,
 data from different $L$'s will march across the $x$ axis
at different rates. The exponent can be determined by tuning $y_m$ to collapse
the data onto a single curve. A picture from my work is shown in Fig.~\ref{fig:alltest}.

Back to Hasenbusch. This is a really professional calculation! There are two big issues he had to address:

The simulations need large volumes, and high statistics. The problem is critical slowing down; 
any observable has a simulation autocorrelation time $\tau$ which scales as $L^p$. The exponent $p$ is around 2
for Metropolis. However, for spin models there are cluster algorithms, where a whole patch of aligned spins
are flipped at once. (Alas, such algorithms do not exist for four-dimensional gauge theories.)
This pushes $p$ down to 0.3-0.4. It makes simulation on volumes as big as $360^3$ sites possible.

\begin{figure}
\begin{center}
\includegraphics[width=\columnwidth,clip]{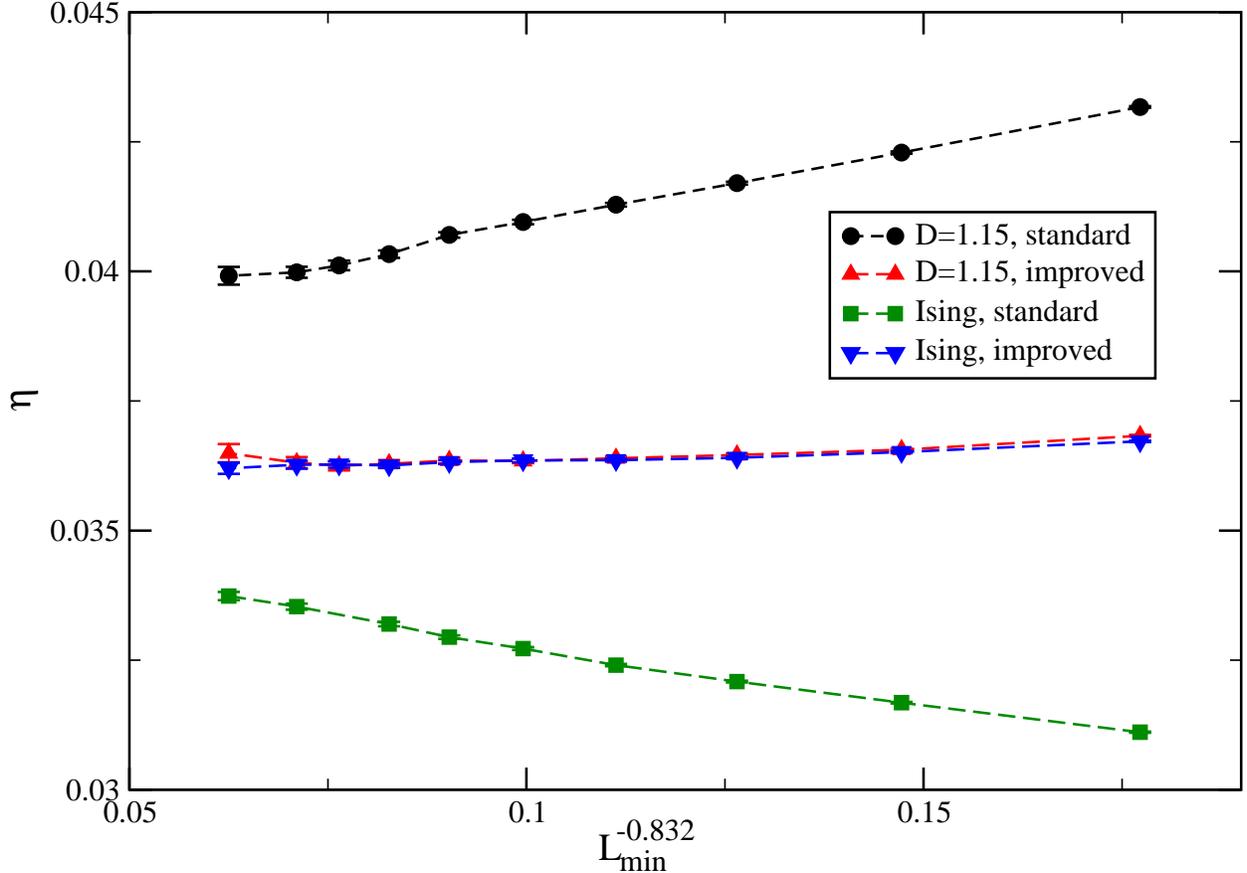}
\caption{
Results for the critical exponent
$\eta$ 
from Ref.~{\protect{\cite{Hasenbusch:2011yya}}}.
obtained by fitting the standard and the improved
magnetic susceptibility for the Ising model and the
Blume-Capel model at $D=1.15$ using Eq.~\protect{\ref{eq:chifit}} with $b=0$.
$L_{min}$ is the minimal lattice size that is taken into
account.
The dashed lines only guide the eye.
}
\label{fig:hasfig1}
\end{center}
\end{figure}

The second, and major, issue is dealing with non-leading corrections to scaling. These are the $L^{-\omega}$
terms in formulas like the one for the magnetic susceptibility,
\bee
\chi = aL^{2-\eta}\times (1+ b L^{-\omega} + \dots) + B
\label{eq:chifit}
\ee
where $B$ is an analytic background. The non-leading term is much less interesting than $\eta$
but the data is so good that it is inescapable in the fits. This is illustrated in Fig.~\ref{fig:hasfig1}.
This shows results of simple fits of Eq.~\ref{eq:chifit} (with $b=0$) to two models in the Ising universality class.
The two models' values for $\eta$ differ by twenty sigma.

The cure is to use ``improved'' operators, ones for which $b$ is very small. This involves two steps.
First, several models are simulated. Besides the Ising model, with spins $s_i=-\pm 1$,
\bee
H = -\beta \sum_{<xy>} s_x s_y
  - h \sum_x s_x  \;\; ,
\ee
Hasenbusch simulated the Blume-Capel model
\bee
H = -\beta \sum_{<xy>} s_x s_y
  + D \sum_x s_x^2  - h \sum_x s_x \;\;  ,
\ee
where now the spins take values $-1,0,1$.
In the limit $D \rightarrow - \infty$  the ``state"  $s=0$ is completely
suppressed, compared with $s=\pm 1$, and therefore the spin-1/2 Ising model
is recovered.
In  $d\ge 2$  dimensions the model undergoes a continuous phase transition
for $-\infty \le  D   < D_{tri} $ at a $\beta_c$ which depends on $D$.
The transition is in the Ising universality class.
For $D > D_{tri}$ the model undergoes a first order phase transition.
The combination of Ising and Blume-Capel models allows Hasenbusch to write down
combinations of correlation functions for which the $b$ term in Eq.~\ref{eq:chifit} vanishes.
The middle lines in Fig.~\ref{fig:hasfig1} (labeled ``improved'') show the nice agreement.

\begin{figure}
\begin{center}
\includegraphics[width=\columnwidth,clip]{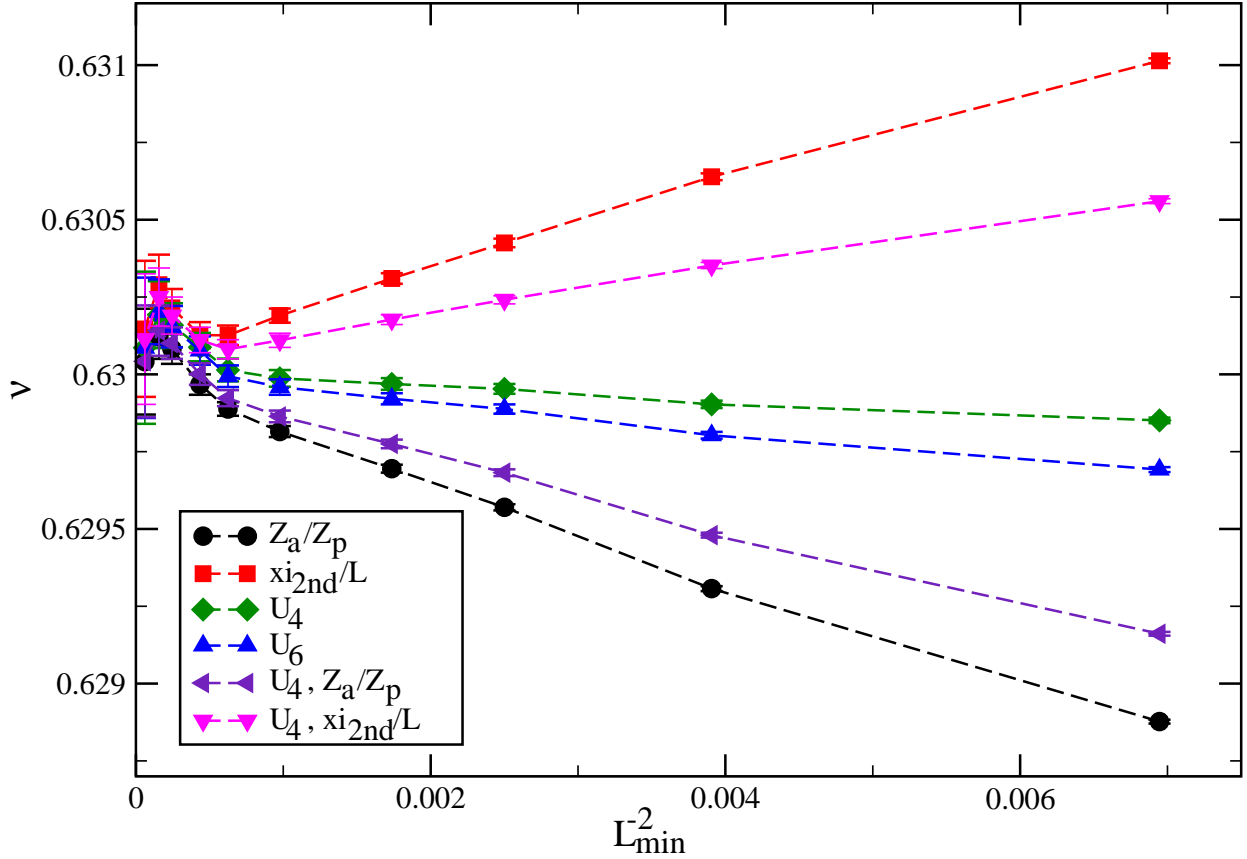}
\caption{
Results 
from Ref.~{\protect{\cite{Hasenbusch:2011yya}}}
for the critical exponent $\nu$ obtained by fitting improved
slopes of various quantities as a function
of $L_{min}^{-2}$, where $L_{min}$ is the minimal lattice
size that is included into the fit. The dashed lines only guide the eye.
}
\label{fig:hasfig4}
\end{center}
\end{figure}

\begin{figure}[htb]
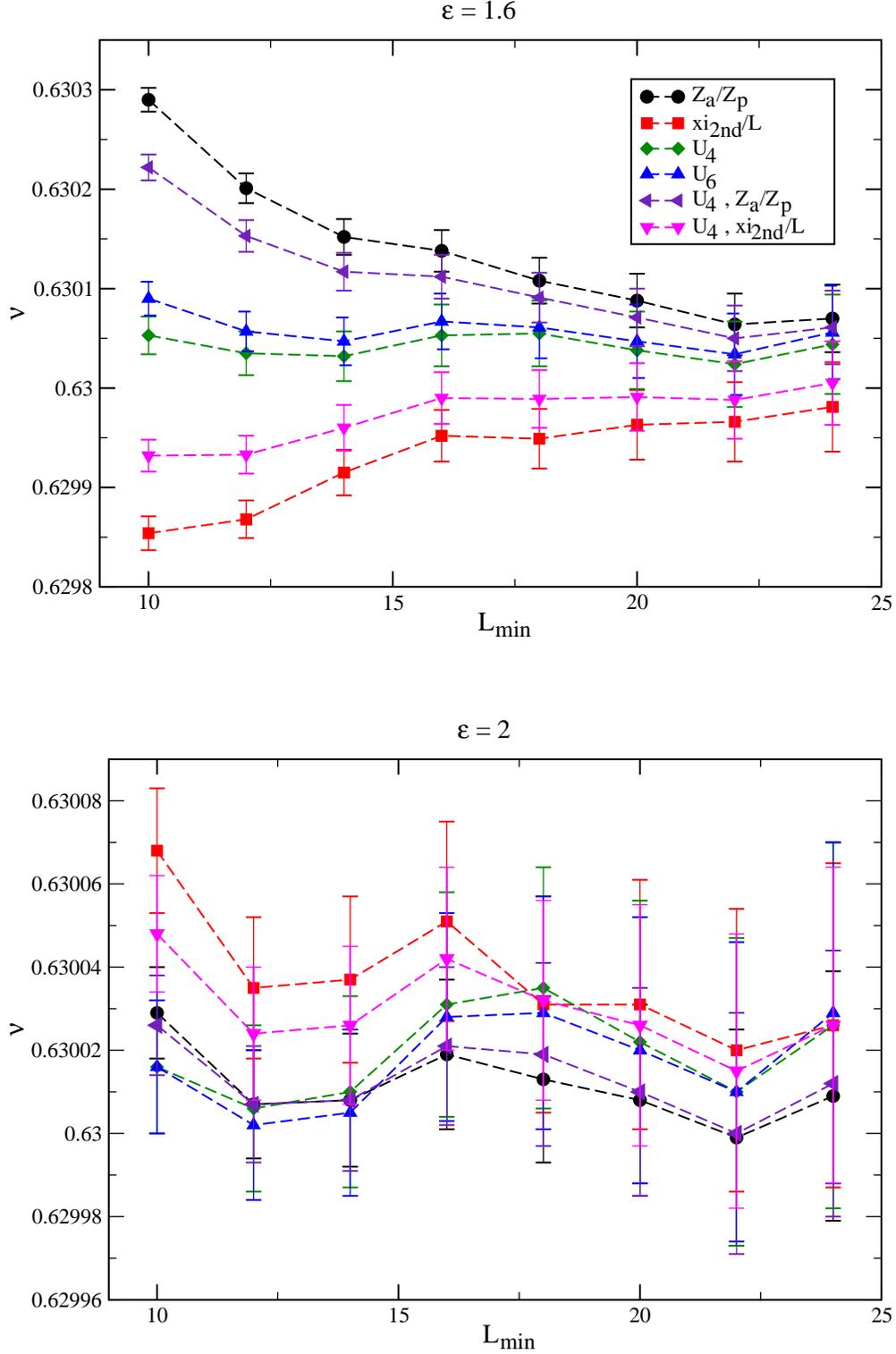

\begin{center}
\includegraphics[width=0.8\columnwidth,clip]{hasfig5.eps}
\vskip1.1cm
\includegraphics[width=0.8\columnwidth,clip]{hasfig6.eps}
\caption{
Results 
from Ref.~{\protect{\cite{Hasenbusch:2011yya}}}
for the critical exponent $\nu$ obtained by fitting
improved slopes of various
couplings at $Z_a/Z_p=0.5425$ as a function of $L_{min}$. In the upper
part of the figure the correction exponent is fixed to $\epsilon=1.6$
and in the lower part it is fixed to  $\epsilon=2$.
The dashed lines should only guide the eye.
}
\label{fig:hasfig5}
\end{center}
\end{figure}

A different set of operators gives $\nu$. Fig.~\ref{fig:hasfig4} and Fig.~\ref{fig:hasfig5}
show results to fits of various improved observables to the functional form
\bee
 S = a(D) L^{1/\nu} \times (1 + b L^{-\epsilon}) 
\ee
with $\epsilon$ to either $1.6$ or $2$.

At the end of the day Hasenbusch had $\nu=0.63002(10)$ and
$\eta=0.03627(10)$. He quotes the best experimental numbers as $\nu=0.632(2)$ and $\eta=0.041(5)$.
That would be a really big box on Fig.~\ref{fig:dds}!

\subsection{QCD}

A long prehistory of nuclear physics without constituents for the proton and neutron
is coming full circle as people
try to compute nuclear properties from lattice QCD simulations. The flavor content of
constituents (the ``eightfold way'' and then fractional charge quarks) came along before
dynamics, and then there were the deep inelastic scattering experiments at SLAC in the late 60's.
Quickly following the discovery of asymptotic freedom by Politzer \cite{Politzer:1973fx} and
Gross and Wilczek \cite{Gross:1973id} came the realization that QCD was the theory of the strong interactions
(compare Ref.~\cite{Fritzsch:1973pi}), but how to get confinement was unknown before Wilson.
The name ``QCD'' is Gell-Mann's.
Wilson and Kogut and Susskind and others did strong coupling calculations in the 70's. Monte Carlo
for pure gauge theories began with Creutz in 1979 and it did not take long before others were attempting
to do Monte Carlo with fermions. These were heroic times with great ideas and inadequate computers.
Most lattice people would say that ``serious'' calculations (meaning, reasonably high precision)
started about 15 years ago. Now lattice QCD is mature and professional.

If you were lattice students, I would start talking about how to do specific calculations
in an efficient way. But that's not a good lecture for this audience. Instead I want to talk about two things:
What is the big picture?  and Is there some simple way to organize what we know about QCD
and its close relatives?
As I am writing this, I am thinking about two Tasi 2017 lecturers, Jim Halverson and Joanna Erdmenger,
and papers they have written about confining systems related to QCD: 
Refs.~\cite{Halverson:2018xge,Erdmenger:2014fxa}.

The qualitative features of QCD are
\begin{enumerate}
\item asymptotic freedom
\item confinement
\item chiral symmetry breaking, when the constituents are light
\end{enumerate}

We use bits of asymptotic freedom in our simulations (taking the bare coupling to zero takes the 
lattice spacing to zero), and some of us measure lattice quantities which give a running coupling as an output,
but we tend not to think in terms of $\Lambda$ parameters (say, to set the overall
scale). They are hard to compute, and they are tied a bit too closely to some perturbative scheme
for our nonperturbative taste.

No first principles analytic calculation exists which allows controlled quantitative prediction
of the consequences of confinement or chiral symmetry breaking in QCD.
There are analytic calculations in models (the Schwinger model, $1+1$ dimensional QED; the `t Hooft
model, $1+1$ dimensional large $N_c$ QCD;  Polyakov's solution of 3-d $U(1)$ gauge theory via
duality transformation \cite{Polyakov:1976fu}, see also \cite{Polyakov:1987ez};
 and of course much more modern duality stories by our organizers) but if you want to
predict (postdict) the mass
of the proton in MeV, right now lattice Monte Carlo is the only game in town.

For us, confinement is a given. We care about its consequences.

And for us, chiral symmetry breaking is also a given. This is actually a pretty big part of lattice QCD.
It is another use of effective field theories. We have two issues: first, the chiral Lagrangian is the
low energy effective field theory of QCD. It has its own set of ``fundamental'' parameters,
like the pion decay constant, and the condensate. We claim to have a more fundamental theory (QCD),
and so we can compute these parameters. The other issue is that most lattice calculations
are not done at the physical values of the up and down quark masses due to the cost of simulating there.
We want to produce results at the physical point to compare to experiment. There is a lot of use
of chiral Lagrangians to do these extrapolations.
(Today there are calculations done at the physical point; some people want to be ``theory-free.'')

All the possibilities  \cite{Peskin:1980gc,Preskill:1980mz,Kosower:1984aw}
about chiral symmetry breaking have been seen in lattice simulations.
 When the fermions make up a complex representation of the gauge group, the expected
pattern of chiral symmetry breaking is $SU(N_f)\times SU(N_f)\rightarrow SU(N_f)$. 
With $N_f$ Dirac fermions (or $2N_f$ Majoranas)
in a real representation of the gauge group, the symmetry breaking pattern is $SU(2N_f)\rightarrow SO(2N_f)$.
With a pseudoreal fermion representation, it is $SU(2N_f)\rightarrow Sp(2N_f)$.
You see the pattern in the spectrum of would-be Goldstone bosons.

The lattice ideology is very clean: we simulate $SU(3)$ gauge theory with $N_f$ flavors of quarks,
at their physical mass values, and what we see is a prediction of QCD without any model dependence.
And it works! Here are two pictures, taken from the Quark Model review of the Particle Data Book \cite{Tanabashi:2018oca}.
Fig~\ref{fig:kronfeld} is a summary of light hadron spectroscopy given to me by Andreas Kronfeld.
Fig.~\ref{fig:davies} is a picture of heavy quark spectroscopy, from Christine Davies.

\begin{figure}
\begin{center}
\includegraphics[width=0.6\textwidth,clip]{spectrum-pdg-2016.eps}
\end{center}
\caption{
Hadron spectrum from lattice QCD.
        Comprehensive results for mesons and baryons are from
        MILC\protect{\cite{Aubin:2004wf,Bazavov:2009bb}},
        PACS-CS\protect{\cite{Aoki:2008sm}},
        BMW\protect{\cite{Durr:2008zz}},
        QCDSF\protect{\cite{Bietenholz:2011qq}}, and
        ETM\protect{\cite{Alexandrou:2014sha}}.
        Results for $\eta$ and $\eta'$ are from
        RBC \& UKQCD\protect{\cite{Christ:2010dd}},
        Hadron Spectrum\protect{\cite{Dudek:2011tt}}
               (also the only $\omega$ mass),
        UKQCD\protect{\cite{Gregory:2011sg}}, and
        Michael, Ottnad, and Urbach\protect{\cite{Michael:2013gka}}.
        Results for heavy-light hadrons from
        Fermilab-MILC\protect{\cite{Bernard:2010fr}},
        HPQCD\protect{\cite{Gregory:2010gm,Dowdall:2012ab}}, and
        Mohler and Woloshyn\protect{\cite{Mohler:2011ke}}.
        Circles, squares, diamonds, and triangles stand for staggered, Wilson,
twisted-mass Wilson, and chiral sea quarks, respectively.
        Asterisks represent anisotropic lattices.
        Open symbols denote the masses used to fix parameters.
        Filled symbols (and asterisks) denote results.
        Red, orange, yellow, green, and blue stand for increasing numbers of
ensembles
        (i.e., lattice spacing and sea quark mass).
        Black symbols stand for results with 2+1+1 flavors of sea quarks.
        Horizontal bars (gray boxes) denote experimentally measured masses
(widths).
        $b$-flavored meson masses are offset by $-4000$~MeV.}
\label{fig:kronfeld}
\end{figure}

\begin{figure}
\begin{center}
\includegraphics[width=0.6\textwidth,clip]{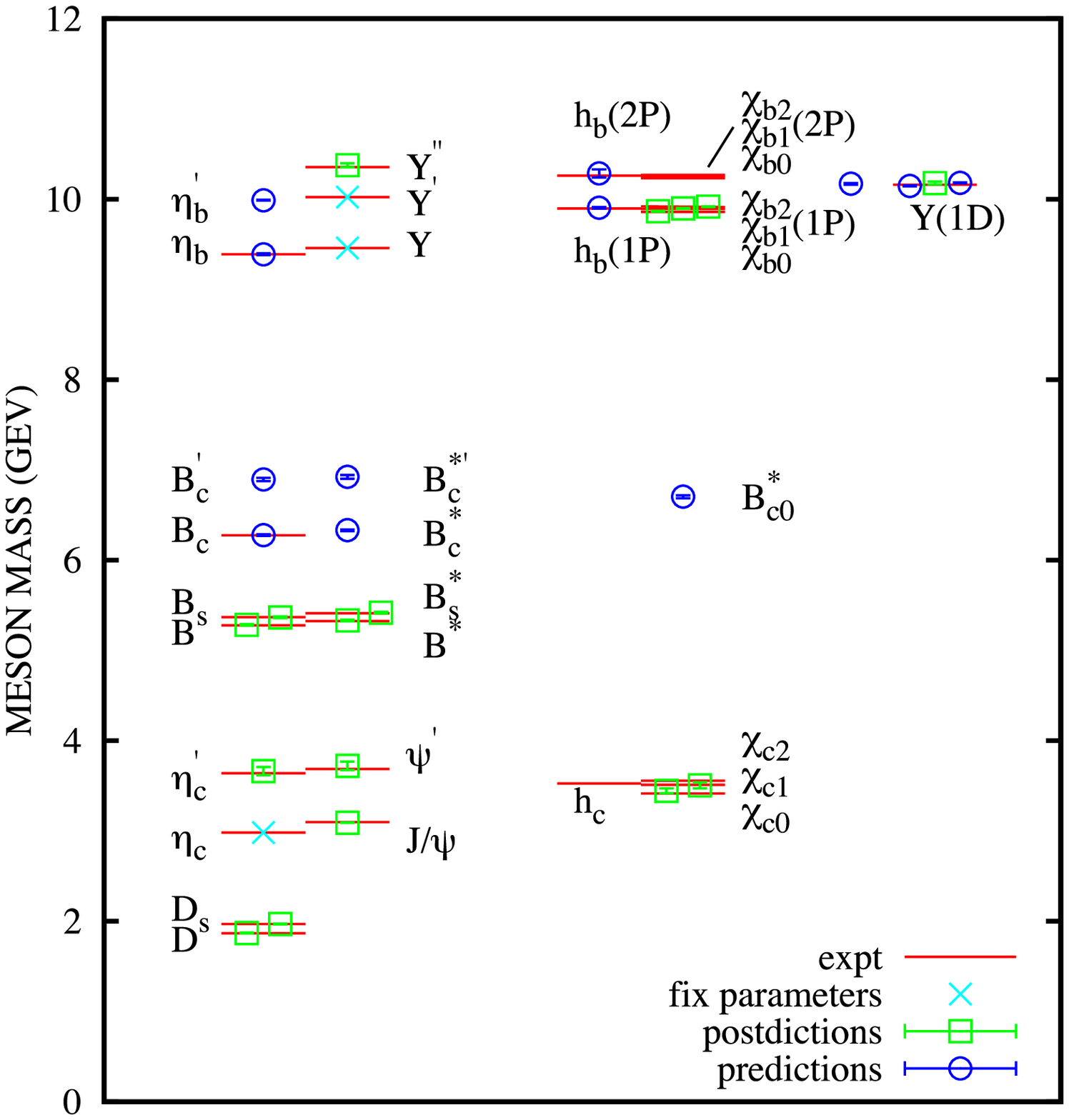}
\end{center}
\caption{
Spectroscopy for mesonic systems containing one or more heavy quarks
(adapted from  \protect{\cite{Dowdall:2012ab}}).
Particles whose masses are used to fix lattice parameters are shown with crosses;
the authors distinguish between ``predictions'' and ``postdictions'' of their calculation.
Lines represent experiment.
}
\label{fig:davies}
\end{figure}

This is a typical story in theoretical physics (thnk about high order QED calculations) but perhaps it is
not completely satisfactory. 
Is there a physical picture for QCD? Yes, in fact, there are several stories:
\begin{itemize}
\item
't Hooft's large $N_c$ limit
\item Heavy quark effective theory
\item Nonrelativistic QCD
\end{itemize}
The stories are not complete. Think in analogy with the Wigner-Eckhart theorem, a matrix element is
a product of a Clebsch-Gordon coefficient and a reduced matrix element. Symmetry considerations
can tell you part of what you want to know (the Clebsches) but not everything (not the reduced
matrix element).

Begin with NRQCD. This is a classic effective field theory story: write down an effective Lagrangian whose
terms are given by an expansion parameterized by the velocity of the quark. Match its coefficients
against some analog calculation in full QCD. Fig.~\ref{fig:davies} is actually based on a lattice version of this 
expansion.

It's simpler, but a bit more uncontrolled, just to think about a Schrodinger equation, heavy quarks moving
in a potential $V(r) \sim -\alpha/r + \sigma r$. This is a really old story. (It goes back to
the discovery of charmonium in 1974.) It works, but the issue is that it does not work completely.
It is very hard to understand fine structure. Is the confining potential the fourth component of a
four vector, or is it a scalar, or something else?

Heavy quark effective theory is similar. The expansion is in the inverse mass. This is used for
heavy-light systems. If the heavy quark in a meson is sufficiently heavy, the light quark (or any other light degree
of freedom) doesn't care. The ``Clebsch'' expansion is
\bee
\svev{O(M)} = \sum_p \frac{\Lambda^p}{M^p}
\ee
where the powers $p$ can be motivated by heavy quark effective theory, but some calculation with dynamics
(like the lattice) is needed to give the $\Lambda$'s.

The summer school had a lecture series on large $N$ by Maldacena. The two-index version,
't Hooft's ~\cite{'tHooft:1973jz} large-$N_c$ limit, is the one for QCD.
Remember the ingredients: graph counting using 
the Fierz identity
\bee
(t^a)^i_j(t^a)^k_l= \frac{1}{2}[\delta^i_l \delta^k_j - \frac{1}{N_C} \delta^i_j\delta^k_l],
\label{eq:fierz}
\ee
the `t Hooft coupling $\lambda=g^2 N_c$,
and the beta function
\bee
\mu \frac{dg^2}{d\mu} = (\frac{11}{3}N_C - \frac{2}{3}N_F)g^4 \rightarrow  \frac{d\lambda}{d\mu}=\frac{11}{3}\lambda^2+\dots
\ee
suggest matching across $N_c$ by comparing at fixed $\lambda$ but $N_c$ taken to infinity.
This says $g\sim 1/\sqrt{N_c}$, and in Feynman graphs, Eq.~\ref{eq:fierz} says
a gluon line basically behaves like
two fermion lines. Graph weight is color weight. 
(Another classic paper: Witten's Ref.~\cite{Witten:1979kh}.)
There is a famous large-$N_c$ phenomenology, which mostly works: Mesons are $\bar q q$ pairs,
which are narrow ($\Gamma/M\sim 1/N_c$),  meson masses vary with quark mass in a nearly $N_c$ independent
way, and so on.  The ``Clebsch formula'' 
is
\bee
\svev{O(N_c)} = \sum_p N_c^p c_p
\ee
where the $c_p$'s are outside the model.

Graph counting -- this seems naive, but it works. Here are a few pictures.

\begin{figure}
\begin{center}
\includegraphics[width=0.5\textwidth,clip]{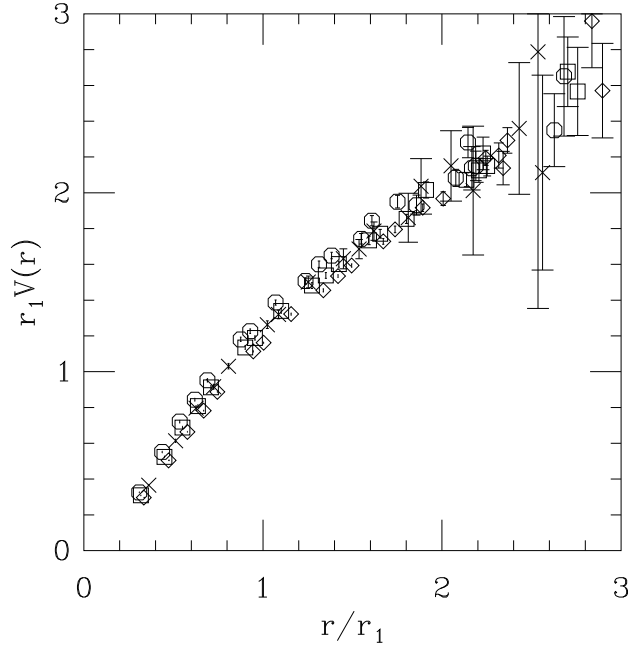}
\end{center}
\caption{Comparison of the dimensionless combination $r_1V(r)$ vs $r/r_1$ from data sets matched in
quark mass, at $(m_{PS}/m_V)^2=0.4$. Symbols are crosses for $N_c=2$, octagons for $N_c=3$,
squares for $N_c=4$ and diamonds for $N_c=5$.
\label{fig:manyrvr}}
\end{figure}

Fig.~\ref{fig:manyrvr} is a plot of the static potential, computed from Wilson loops.
This is my own data \cite{DeGrand:2016pur}, nothing special, not high precision. 
But it will take us into the next part of the
story. There are several things to explain:
First, this is a picture of many different systems. The plotting symbols are for data at different
values of $N_c$ ranging from 2 to 5. Second, the axis are scaled with a common overall scale
(the place where $r^2 dV(r)/dr= 1.0$). There are three obvious features:
First, $V(r)$ looks linear at long distance: that is linear confinement. Second, it looks
Coulombic at short distance; that is what you would expect thinking naively about asymptotic freedom;
one gluon exchange. Third, the potentials for all the different $N_c$'s look identical.

There is a good review of lattice large $N_c$ tests of pure gauge theory by Lucini and Panero \cite{Lucini:2012gg}.
 The literature
on large $N_c$ with fermions is  small.  Nearly all studies are done in quenched approximation,
neglecting virtual quark anti-quark pairs.
The most extensive study of meson spectroscopy and matrix elements
is done by Bali et al \cite{Bali:2013kia}.
They cover $N_c=2-7$ and 17.

The results of all these studies are easy to state:
large $N_c$ counting works very well. (When I told Phillippe de Forcrand I was doing
large $N_c$ on the lattice, he said, ``Why are you working on things
where you already know the answer?'') Specifically,
\begin{itemize}
\item Matching lattice spacings across $N_c$ basically happens when the bare 't Hooft couplings
$\lambda=g^2 N_c$ are matched
\item Dimensionless ratios of masses in the pure gauge systems are nearly $N_c$ independent
\item The dependence of meson masses on quark mass is nearly $N_c$ independent
\item Meson matrix elements, like the pseudoscalar decay constant, scale like $\sqrt{N_c}$.
Other matrix elements scale the way they are supposed to (see Ref.~\cite{Donini:2016lwz}).
That is, graph-counting gives a characteristic $N_c^p$ dependence on any process.
\end{itemize}

\begin{figure}
\begin{center}
\includegraphics[width=0.7\textwidth,clip]{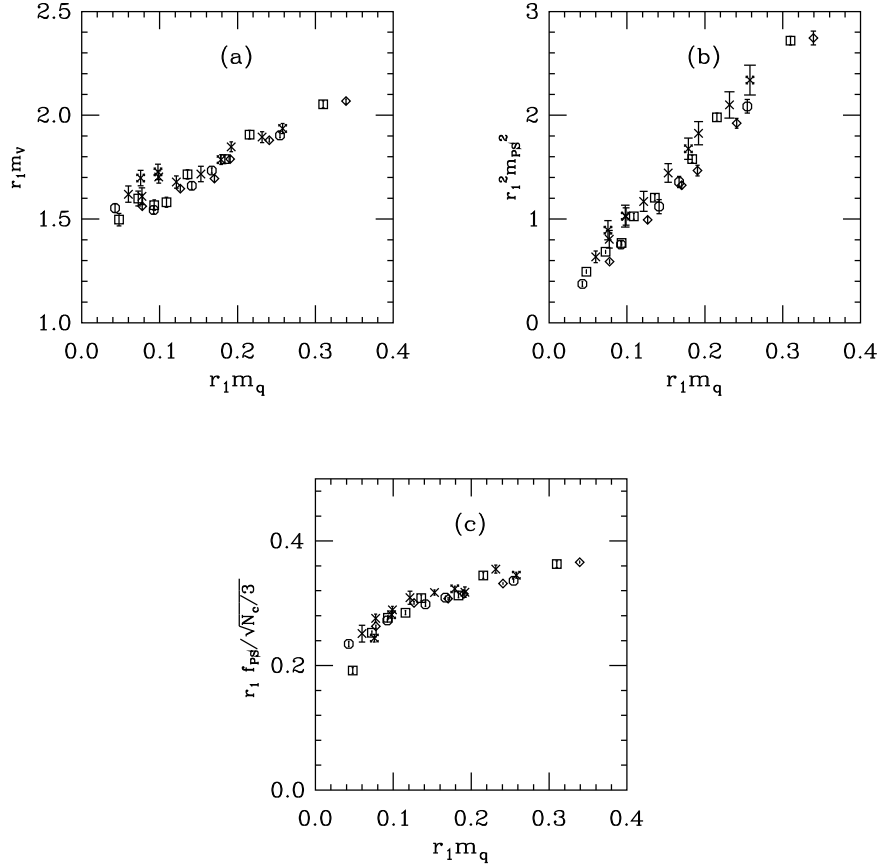}
\end{center}
\caption{ (a) Vector meson mass versus quark mass.
(b) Squared pseudoscalar mass versus quark mass. 
(c) Pseudoscalar decay constant divided by $\sqrt{N_c/3}$ so that curve collapse signals the
correct large $N_c$ scaling behavior,
 versus quark mass.
 Data are crosses for $SU(2)$, squares for $SU(3)$,
octagons for $SU(4)$, and diamonds for $SU(5)$.
\label{fig:multiNc}}
\end{figure}

Fig.~\ref{fig:multiNc} illustrates a few of these points. (Incidentally, panel (b),
with $m_{PS}^2 \propto m_q$, shows that the pion is a Goldstone boson.)

\begin{figure}
\begin{center}
\includegraphics[width=0.6\textwidth,clip]{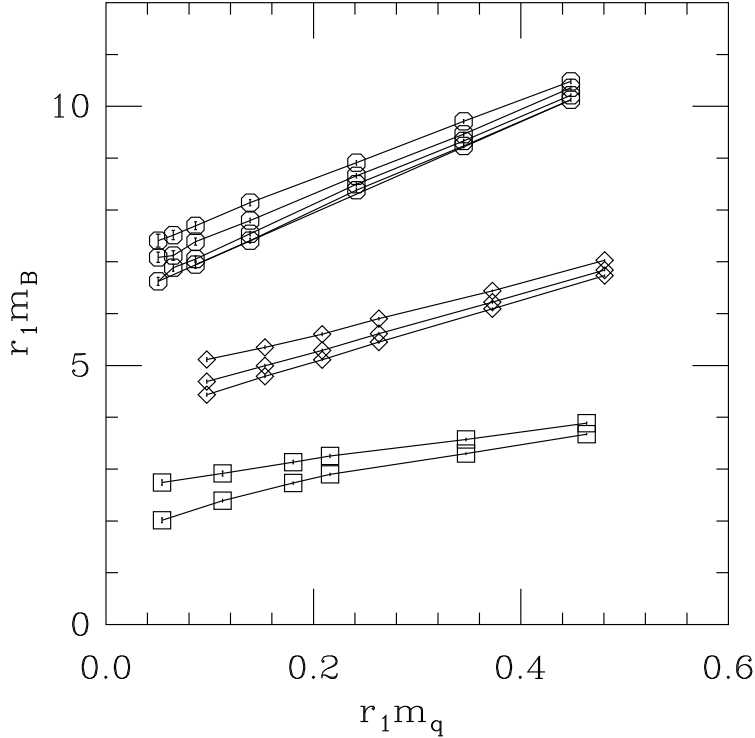}
\end{center}
\caption{Baryon spectroscopy  in units of $r_1$ (0.3 fm in our world)
from $N_c=3$, 5 and 7, plotted as squares, diamonds and octagons, respectively. 
For all $N_c$, higher $J$ lies higher in mass. From {\protect\cite{DeGrand:2012hd}}, mostly.
\label{fig:baryons}}
\end{figure}

Baryons deserve a fuller paragraph. 
Baryons in large $N_c$ can be regarded as many-quark states \cite{Witten:1979kh}
 or as topological objects
in effective theories of mesons\cite{Witten:1983tx,Adkins:1983ya,Gervais:1983wq,Gervais:1984rc}.
Large-$N_c$ mass formulas for baryons have been devised, see
Refs.~\cite{ Jenkins:1993zu,Dashen:1993jt,Dashen:1994qi,Jenkins:1995td,Dai:1995zg,Cherman:2012eg}.
Lattice baryons are discussed in
Ref.~\cite{Jenkins:2009wv} and in my papers
\cite{DeGrand:2012hd,DeGrand:2013nna,Cordon:2014sda}, with $N_c=3$ to 7.
$N$-color baryons with flavor $SU(2)$ symmetry come in isospin-spin locked multiplets, with isospin $I$
 and angular momentum $J$ locked and equal to
$I=J=1/2$, $3/2,\dots,N_c/2$. The spectrum is that of a rigid rotor,
 \bee
 M(N,J) = N_c m_0  + \frac{J(J+1)}{N_c}B +\dots .
\ee
The two parameters $m_0$ and $B$ are ``typical QCD sizes'' of a few hundred MeV, but
their quark mass dependence is different. $m_0$ grows with quark mass while $B$ falls.
The $N_c m_0$ term is the ``derivation of the quark model'' from large $N_c$; at leading
order a baryon acts like it is made of $N_c$ quarks.
Look at Fig.~\ref{fig:baryons}.

This doesn't quite get us back to Fig.~\ref{fig:kronfeld} so let's look at it more closely.
The lightest states are the pions, they are pseudo Goldstone bosons, pseudo because the
up and down quarks are not quite massless. The kaons have a massive strange quark so they are
heavier. Lattice people can vary the quark mass at will and see this. The $\eta$ and $\eta'$ mesons
(noticed by a sharp eyed student) are, appropriately for this Tasi, heavy because of the axial
anomaly.

The rest of the picture can (mostly) be understood with the quark model,
and the easiest quark model to write down is the the ``color hyperfine interaction''
 picture of De Rujula, Georgi and Glashow \cite{DeRujula:1975qlm}.
The Hamiltonian for a bound state of $N_i$ quarks of constituent mass $m_i$ and spin $\vec S_i$,
all in identical S-wave spatial wave functions, is taken to be
\bee
H = \sum_i N_i m_i + \sum_{i\ne j} C_{ij}\vec S_i\cdot \vec S_j.
\ee
$m_i$ is a function of the actual physical quark mass.
$C_{ij}$ is a mass-dependent constant representing a magnetic hyperfine interaction between the quarks.
Think of it as a product of magnetic moments, so that it is smaller for heavier quarks.

This will give you a plausible story for all the states except the Goldstones. Vector mesons
are heavier than the pseudoscalars because $S_i\cdot \vec S_j$ is 1/4, not -3/4.

If you include the color factor in the hyperfine interaction, you can account for the large $N_c$ counting
of spin splittings.
This is simple: one gluon exchange in a meson is 
\bee
V_m= g^2(\frac{1}{\sqrt{N_c}})^2(t^A)^a_b(t^A)^c_d \delta^d_a \delta^b_c = g^2\frac{N_c^2-1}{2N_c}
\ee
(the leading square roots are the color singlet factors)
while for baryons (minus sign for quark vs antiquark, antisymmetric color wave function)
\bee
V_B= -g^2(\frac{1}{\sqrt{N_c!}})^2 \epsilon_{abc\dots}\epsilon^{a'b'c'\dots} (t^A)^a_{a'}(t^A)^b_{b'} \delta^c_{c'} \delta^d_{d'}
= g^2\frac{N_c+1}{N_c}
\ee
The factor of $g^2$  is $1/N_c$ times the 't Hooft
coupling $\lambda$. So for mesons, the color factor is basically $\lambda$ and for baryons it is
$\lambda/N_c$.

Finally, I have to say two things: First, the models won't get you the reduced matrix elements.
 And second, about the only model input lattice people use
in a real simulation is that  the interpolating fields for hadrons are basically quark model
wave functions. It  is still true that a lattice QCD calculation starts with the
 Lagrangian for QCD plus irrelevant operators, and (once the appropriate limits are taken) lattice
QCD predictions are QCD predictions without the ``lattice'' adjective.

As I said, most lattice QCD people are doing complicated calculations of matrix elements.
 The motivation is either to
test the standard model (most processes happen inside hadrons)
or to calculate in QCD (because it is interesting for itself). 
To reference this, I will just cite
a couple of white papers \cite{Cirigliano:2019jig,Lehner:2019wvv}
which I enjoyed reading as overviews (at a more technical level than I am writing).

Perhaps this is enough about QCD-like systems.
As the number of fermionic degrees of freedom rises, physics changes qualitatively. It is thought 
that there is
a crossover from  confinement to ``infrared conformality,'' which is what we call a system
which is conformal at long distances. One could imagine that this could happen if the
 the beta function had a second zero, in addition to the Gaussian one, which was infrared attractive.
($\beta(g)= - c_1 g^3 + c_2 g^5$). This is called a ``Banks-Zaks fixed point'' \cite{Banks:1981nn}.
Most of Caswell's  paper on the two loop beta function \cite{Caswell:1974gg} is about this physics.

People try to guesstimate when this happens by playing with the perturbative two-loop
beta function. At the bottom of the conformal window, or with the number of fermionic degrees of freedom
barely above the crossover value, this would be an untrustworthy calculation because
any zero would occur deep in strong coupling.
The program to look for zeroes of beta functions and then to measure critical indices
was pretty big from about 2008 to 2013 or so. It is a tough game.
People still argue about where the bottom of the window is; 
the best guess I can give is that it is someplace between $N_f=8$  and 12 for $SU(3)$ with fundamental fermions.

At very large $N_f$ asymptotic freedom is lost (of course). I don't know of much lattice work up here.


\section{Instead of a conclusion}

Let's return to a big picture summary. Most of this year's Tasi was 
about idealized systems with special properties, 
such as supersymmetry or conformality.
 Lattice Monte Carlo is a generic set of techniques for studying strongly interacting
Euclidean space quantum field theories and equilibrium statistical systems.
 The weaknesses of lattice methods are
\begin{itemize}
\item If your system has special properties, you may not be able to exploit
 these properties in a lattice
calculation.
\item Things you ignored at the start may come back to haunt you.
 The biggest of these issues for systems tuned to be critical or nearly so is that the cost of studying
a system of size $L$ in $D$ dimensions scales at least like $L^D$.
\end{itemize}
The strengths of lattice techniques are
\begin{itemize}
\item
Your system does not have to be (too) special.
\item
Sometimes, you can compute nonperturbative observables in a controlled way.
\end{itemize}



\begin{acknowledgments}
I was surprised to be asked to lecture at this Tasi, and I hope I said useful things to 
the students and
to their scientific community. Thanks to Will Jay for reading an earlier version of these notes.
This work was supported by the US Department of Energy under grant DE-SC0010005.
Tasi, or the ``Theoretical Advanced Study Institute,'' is supported
by the National Science Foundation with award 1819655.
\end{acknowledgments}



\begin{thebibliography}{99}

\bibitem{DeGrand:2006zz} 
  T.~DeGrand and C.~E.~Detar,
  ``Lattice methods for quantum chromodynamics,''
  New Jersey, USA: World Scientific (2006) 345 p

\bibitem{Gattringer:2010zz} 
  C.~Gattringer and C.~B.~Lang,
  ``Quantum chromodynamics on the lattice,''
  Lect.\ Notes Phys.\  {\bf 788}, 1 (2010).
  doi:10.1007/978-3-642-01850-3

\bibitem{Knechtli:2017sna} 
  F.~Knechtli, M.~G\"unther and M.~Peardon,
  ``Lattice Quantum Chromodynamics: Practical Essentials,''
  doi:10.1007/978-94-024-0999-4

\bibitem{Hanada:2018fnp} 
  M.~Hanada,
  ``Markov Chain Monte Carlo for Dummies,''
  arXiv:1808.08490 [hep-th].

\bibitem{Wilson:1974sk} 
  K.~G.~Wilson,
  Phys.\ Rev.\ D {\bf 10}, 2445 (1974).
  doi:10.1103/PhysRevD.10.2445


\bibitem{Drouffe:1978dn} 
  J.~M.~Drouffe and C.~Itzykson,
  Phys.\ Rept.\  {\bf 38}, 133 (1978).
  doi:10.1016/0370-1573(78)90154-0


\bibitem{Kogut:1974ag} 
  J.~B.~Kogut and L.~Susskind,
  Phys.\ Rev.\ D {\bf 11}, 395 (1975).
  doi:10.1103/PhysRevD.11.395


\bibitem{Krasnitz:1998ns} 
  A.~Krasnitz and R.~Venugopalan,
  Nucl.\ Phys.\ B {\bf 557}, 237 (1999)
  doi:10.1016/S0550-3213(99)00366-1
  [hep-ph/9809433].


\bibitem{Wilson:1973jj} 
  K.~G.~Wilson and J.~B.~Kogut,
  Phys.\ Rept.\  {\bf 12}, 75 (1974).
  doi:10.1016/0370-1573(74)90023-4


\bibitem{Creutz:1980zw} 
  M.~Creutz,
  Phys.\ Rev.\ D {\bf 21}, 2308 (1980).
  doi:10.1103/PhysRevD.21.2308


\bibitem{Metropolis:1953am} 
  N.~Metropolis, A.~W.~Rosenbluth, M.~N.~Rosenbluth, A.~H.~Teller and E.~Teller,
  J.\ Chem.\ Phys.\  {\bf 21}, 1087 (1953).
  doi:10.1063/1.1699114

\bibitem{Jansen:2003nt} 
  K.~Jansen,
  Nucl.\ Phys.\ Proc.\ Suppl.\  {\bf 129}, 3 (2004)
  doi:10.1016/S0920-5632(03)02502-7
  [hep-lat/0311039].


\bibitem{DeGrand:2015zxa} 
  T.~DeGrand,
  Rev.\ Mod.\ Phys.\  {\bf 88}, 015001 (2016)
  doi:10.1103/RevModPhys.88.015001
  [arXiv:1510.05018 [hep-ph]].



\bibitem{Catterall:2009it} 
  S.~Catterall, D.~B.~Kaplan and M.~Unsal,
  Phys.\ Rept.\  {\bf 484}, 71 (2009)
  doi:10.1016/j.physrep.2009.09.001
  [arXiv:0903.4881 [hep-lat]].


\bibitem{Fleming:2000fa} 
  G.~T.~Fleming, J.~B.~Kogut and P.~M.~Vranas,
  Phys.\ Rev.\ D {\bf 64}, 034510 (2001)
  doi:10.1103/PhysRevD.64.034510
  [hep-lat/0008009].

\bibitem{Giedt:2008xm} 
  J.~Giedt, R.~Brower, S.~Catterall, G.~T.~Fleming and P.~Vranas,
  Phys.\ Rev.\ D {\bf 79}, 025015 (2009)
  doi:10.1103/PhysRevD.79.025015
  [arXiv:0810.5746 [hep-lat]].


\bibitem{Endres:2009yp} 
  M.~G.~Endres,
  Phys.\ Rev.\ D {\bf 79}, 094503 (2009)
  doi:10.1103/PhysRevD.79.094503
  [arXiv:0902.4267 [hep-lat]].

\bibitem{Kim:2011fw} 
  S.~W.~Kim {\it et al.} [JLQCD Collaboration],
  PoS LATTICE {\bf 2011}, 069 (2011)
  doi:10.22323/1.139.0069
  [arXiv:1111.2180 [hep-lat]].



\bibitem{Bergner:2013nwa} 
  G.~Bergner, I.~Montvay, G.~M\"unster, U.~D.~\"Ozugurel and D.~Sandbrink,
  JHEP {\bf 1311}, 061 (2013)
  doi:10.1007/JHEP11(2013)061
  [arXiv:1304.2168 [hep-lat]].

\bibitem{Ali:2019agk} 
  S.~Ali, G.~Bergner, H.~Gerber, I.~Montvay, G.~M\"unster, S.~Piemonte and P.~Scior,
  arXiv:1902.11127 [hep-lat].


\bibitem{Catterall:2014vka} 
  S.~Catterall, D.~Schaich, P.~H.~Damgaard, T.~DeGrand and J.~Giedt,
  Phys.\ Rev.\ D {\bf 90}, no. 6, 065013 (2014)
  doi:10.1103/PhysRevD.90.065013
  [arXiv:1405.0644 [hep-lat]].

\bibitem{Catterall:2012yq} 
  S.~Catterall, P.~H.~Damgaard, T.~Degrand, R.~Galvez and D.~Mehta,
  JHEP {\bf 1211}, 072 (2012)
  doi:10.1007/JHEP11(2012)072
  [arXiv:1209.5285 [hep-lat]].

\bibitem{Schaich:2018mmv} 
  D.~Schaich,
  arXiv:1810.09282 [hep-lat].


\bibitem{Honda:2013nfa} 
  M.~Honda, G.~Ishiki, S.~W.~Kim, J.~Nishimura and A.~Tsuchiya,
  JHEP {\bf 1311}, 200 (2013)
  doi:10.1007/JHEP11(2013)200
  [arXiv:1308.3525 [hep-th]].

\bibitem{Hanada:2013rga} 
  M.~Hanada, Y.~Hyakutake, G.~Ishiki and J.~Nishimura,
  Science {\bf 344}, 882 (2014)
  doi:10.1126/science.1250122
  [arXiv:1311.5607 [hep-th]].

\bibitem{Honda:2011qk} 
  M.~Honda, G.~Ishiki, J.~Nishimura and A.~Tsuchiya,
  PoS LATTICE {\bf 2011}, 244 (2011)
  doi:10.22323/1.139.0244
  [arXiv:1112.4274 [hep-lat]].

\bibitem{Ishiki:2009sg} 
  G.~Ishiki, S.~W.~Kim, J.~Nishimura and A.~Tsuchiya,
  JHEP {\bf 0909}, 029 (2009)
  doi:10.1088/1126-6708/2009/09/029
  [arXiv:0907.1488 [hep-th]].

\bibitem{Ishii:2008ib} 
  T.~Ishii, G.~Ishiki, S.~Shimasaki and A.~Tsuchiya,
  Phys.\ Rev.\ D {\bf 78}, 106001 (2008)
  doi:10.1103/PhysRevD.78.106001
  [arXiv:0807.2352 [hep-th]].

\bibitem{Ishiki:2008te} 
  G.~Ishiki, S.~W.~Kim, J.~Nishimura and A.~Tsuchiya,
  Phys.\ Rev.\ Lett.\  {\bf 102}, 111601 (2009)
  doi:10.1103/PhysRevLett.102.111601
  [arXiv:0810.2884 [hep-th]].


\bibitem{Shankar}
R.~Shankar, ``Topological insulators -- a review,'' arXiv:1804.06471

\bibitem{Tong:2016kpv} 
  D.~Tong,
  ``Lectures on the Quantum Hall Effect,''
  arXiv:1606.06687 [hep-th].


\bibitem{Nielsen:1980rz} 
  H.~B.~Nielsen and M.~Ninomiya,
  Nucl.\ Phys.\ B {\bf 185}, 20 (1981)
  Erratum: [Nucl.\ Phys.\ B {\bf 195}, 541 (1982)].
  doi:10.1016/0550-3213(81)90361-8, 10.1016/0550-3213(82)90011-6

\bibitem{Nielsen:1981xu} 
  H.~B.~Nielsen and M.~Ninomiya,
  Nucl.\ Phys.\ B {\bf 193}, 173 (1981).
  doi:10.1016/0550-3213(81)90524-1

\bibitem{Jackiw:1975fn} 
  R.~Jackiw and C.~Rebbi,
  Phys.\ Rev.\ D {\bf 13}, 3398 (1976).
  doi:10.1103/PhysRevD.13.3398

\bibitem{Kaplan:1992bt} 
  D.~B.~Kaplan,
  Phys.\ Lett.\ B {\bf 288}, 342 (1992)
  doi:10.1016/0370-2693(92)91112-M
  [hep-lat/9206013].

\bibitem{Shamir:1993zy} 
  Y.~Shamir,
  Nucl.\ Phys.\ B {\bf 406}, 90 (1993)
  doi:10.1016/0550-3213(93)90162-I
  [hep-lat/9303005].


\bibitem{Callan:1984sa} 
  C.~G.~Callan, Jr. and J.~A.~Harvey,
  Nucl.\ Phys.\ B {\bf 250}, 427 (1985).
  doi:10.1016/0550-3213(85)90489-4


\bibitem{Golterman:1992ub} 
  M.~F.~L.~Golterman, K.~Jansen and D.~B.~Kaplan,
  Phys.\ Lett.\ B {\bf 301}, 219 (1993)
  doi:10.1016/0370-2693(93)90692-B
  [hep-lat/9209003].

\bibitem{Kaplan:1995pe} 
  D.~B.~Kaplan and M.~Schmaltz,
  Phys.\ Lett.\ B {\bf 368}, 44 (1996)
  doi:10.1016/0370-2693(95)01485-3
  [hep-th/9510197].


\bibitem{Ginsparg:1981bj}
  P.~H.~Ginsparg and K.~G.~Wilson,
  Phys.\ Rev.\ D {\bf 25}, 2649 (1982).
  doi:10.1103/PhysRevD.25.2649


\bibitem{Neuberger:1997fp}
  H.~Neuberger,
  Phys.\ Lett.\ B {\bf 417}, 141 (1998)
  doi:10.1016/S0370-2693(97)01368-3
  [hep-lat/9707022].



\bibitem{Luscher:2000hn} 
  M.~Luscher,
  Subnucl.\ Ser.\  {\bf 38}, 41 (2002)
  doi:10.1142/9789812778253\_0002
  [hep-th/0102028].

\bibitem{Brower:2012vk} 
  R.~C.~Brower, H.~Neff and K.~Orginos,
  Comput.\ Phys.\ Commun.\  {\bf 220}, 1 (2017)
  doi:10.1016/j.cpc.2017.01.024
  [arXiv:1206.5214 [hep-lat]].



\bibitem{Golterman:2000hr} 
  M.~Golterman,
  Nucl.\ Phys.\ Proc.\ Suppl.\  {\bf 94}, 189 (2001)
  doi:10.1016/S0920-5632(01)00953-7
  [hep-lat/0011027].

\bibitem{Golterman:2004qv} 
  M.~Golterman and Y.~Shamir,
  Phys.\ Rev.\ D {\bf 70}, 094506 (2004)
  doi:10.1103/PhysRevD.70.094506
  [hep-lat/0404011].


\bibitem{Eichten:1985ft} 
  E.~Eichten and J.~Preskill,
  Nucl.\ Phys.\ B {\bf 268}, 179 (1986).
  doi:10.1016/0550-3213(86)90207-5

\bibitem{Golterman:1992yha} 
  M.~F.~L.~Golterman, D.~N.~Petcher and E.~Rivas,
  Nucl.\ Phys.\ B {\bf 395}, 596 (1993)
  doi:10.1016/0550-3213(93)90049-U
  [hep-lat/9206010].

\bibitem{Luscher:2000zd} 
  M.~Luscher,
  JHEP {\bf 0006}, 028 (2000)
  doi:10.1088/1126-6708/2000/06/028
  [hep-lat/0006014].


\bibitem{Grabowska:2016bis} 
  D.~M.~Grabowska and D.~B.~Kaplan,
  Phys.\ Rev.\ D {\bf 94}, no. 11, 114504 (2016)
  doi:10.1103/PhysRevD.94.114504
  [arXiv:1610.02151 [hep-lat]].

\bibitem{Grabowska:2015qpk} 
  D.~M.~Grabowska and D.~B.~Kaplan,
  Phys.\ Rev.\ Lett.\  {\bf 116}, no. 21, 211602 (2016)
  doi:10.1103/PhysRevLett.116.211602
  [arXiv:1511.03649 [hep-lat]].




\bibitem{Simmons-Duffin:2016gjk} 
  D.~Simmons-Duffin,
  ``The Conformal Bootstrap,'' in the Proceedings of TASI - 2015, New Jersey, USA: World Scientific (2016);
  doi:10.1142/9789813149441\_0001  [arXiv:1602.07982 [hep-th]].

\bibitem{Simmons-Duffin:2015qma} 
  D.~Simmons-Duffin,
  JHEP {\bf 1506}, 174 (2015)
  doi:10.1007/JHEP06(2015)174
  [arXiv:1502.02033 [hep-th]].


\bibitem{Hasenbusch:2011yya} 
  M.~Hasenbusch,
  Phys.\ Rev.\ B {\bf 82}, 174433 (2010)
  doi:10.1103/PhysRevB.82.174433
  [arXiv:1004.4486 [cond-mat.stat-mech]].


\bibitem{Cardy:1996xt}
  J.~L.~Cardy,
  ``Scaling And Renormalization In Statistical Physics,''
{\it  Cambridge, UK: Univ. Pr. (1996). (Cambridge lecture notes in physics: 3)}

\bibitem{Metlitski:2015eka} 
  M.~A.~Metlitski and A.~Vishwanath,
  Phys.\ Rev.\ B {\bf 93}, no. 24, 245151 (2016)
  doi:10.1103/PhysRevB.93.245151
  [arXiv:1505.05142 [cond-mat.str-el]].



\bibitem{Kajantie:1997vc} 
  K.~Kajantie, M.~Karjalainen, M.~Laine and J.~Peisa,
  Phys.\ Rev.\ B {\bf 57}, 3011 (1998)
  doi:10.1103/PhysRevB.57.3011
  [cond-mat/9704056].


\bibitem{Fisher:1982xt} 
  M.~E.~Fisher and A.~N.~Berker,
  Phys.\ Rev.\ B {\bf 26}, 2507 (1982).
  doi:10.1103/PhysRevB.26.2507



\bibitem{Cheng:2013xha} 
  A.~Cheng, A.~Hasenfratz, Y.~Liu, G.~Petropoulos and D.~Schaich,
  Phys.\ Rev.\ D {\bf 90}, no. 1, 014509 (2014)
  doi:10.1103/PhysRevD.90.014509
  [arXiv:1401.0195 [hep-lat]].


\bibitem{DeGrand:2009hu} 
  T.~DeGrand,
  Phys.\ Rev.\ D {\bf 80}, no. 11, 114507 (2009)
  doi:10.1103/PhysRevD.80.114507
  [arXiv:0910.3072 [hep-lat]].










\bibitem{Politzer:1973fx} 
  H.~D.~Politzer,
  Phys.\ Rev.\ Lett.\  {\bf 30}, 1346 (1973).
  doi:10.1103/PhysRevLett.30.1346

\bibitem{Gross:1973id} 
  D.~J.~Gross and F.~Wilczek,
  Phys.\ Rev.\ Lett.\  {\bf 30}, 1343 (1973).
  doi:10.1103/PhysRevLett.30.1343

\bibitem{Fritzsch:1973pi} 
  H.~Fritzsch, M.~Gell-Mann and H.~Leutwyler,
  Phys.\ Lett.\  {\bf 47B}, 365 (1973).
  doi:10.1016/0370-2693(73)90625-4


\bibitem{Halverson:2018xge} 
  J.~Halverson and P.~Langacker,
  PoS TASI {\bf 2017}, 019 (2018)
  doi:10.22323/1.305.0019
  [arXiv:1801.03503 [hep-th]].


\bibitem{Erdmenger:2014fxa} 
  J.~Erdmenger, N.~Evans and M.~Scott,
  Phys.\ Rev.\ D {\bf 91}, no. 8, 085004 (2015)
  doi:10.1103/PhysRevD.91.085004
  [arXiv:1412.3165 [hep-ph]].

\bibitem{Polyakov:1976fu} 
  A.~M.~Polyakov,
  Nucl.\ Phys.\ B {\bf 120}, 429 (1977).
  doi:10.1016/0550-3213(77)90086-4

\bibitem{Polyakov:1987ez}
  A.~M.~Polyakov,
  ``GAUGE FIELDS AND STRINGS,''
{\it  CHUR, SWITZERLAND: HARWOOD (1987) 301 P. (CONTEMPORARY CONCEPTS IN PHYSICS, 3)}



\bibitem{Peskin:1980gc} 
  M.~E.~Peskin,
  Nucl.\ Phys.\ B {\bf 175}, 197 (1980).
  doi:10.1016/0550-3213(80)90051-6

\bibitem{Preskill:1980mz} 
  J.~Preskill,
  Nucl.\ Phys.\ B {\bf 177}, 21 (1981).
  doi:10.1016/0550-3213(81)90265-0

\bibitem{Kosower:1984aw} 
  D.~A.~Kosower,
  Phys.\ Lett.\  {\bf 144B}, 215 (1984).
  doi:10.1016/0370-2693(84)91806-9



\bibitem{Tanabashi:2018oca} 
  M.~Tanabashi {\it et al.} [Particle Data Group],
  Phys.\ Rev.\ D {\bf 98}, no. 3, 030001 (2018).
  doi:10.1103/PhysRevD.98.030001

\bibitem{Aubin:2004wf} 
  C.~Aubin {\it et al.},
  Phys.\ Rev.\ D {\bf 70}, 094505 (2004)
  doi:10.1103/PhysRevD.70.094505
  [hep-lat/0402030].

\bibitem{Bazavov:2009bb} 
  A.~Bazavov {\it et al.} [MILC Collaboration],
  Rev.\ Mod.\ Phys.\  {\bf 82}, 1349 (2010)
  doi:10.1103/RevModPhys.82.1349
  [arXiv:0903.3598 [hep-lat]].

\bibitem{Aoki:2008sm} 
  S.~Aoki {\it et al.} [PACS-CS Collaboration],
  Phys.\ Rev.\ D {\bf 79}, 034503 (2009)
  doi:10.1103/PhysRevD.79.034503
  [arXiv:0807.1661 [hep-lat]].

\bibitem{Durr:2008zz} 
  S.~Durr {\it et al.},
  Science {\bf 322}, 1224 (2008)
  doi:10.1126/science.1163233
  [arXiv:0906.3599 [hep-lat]].

\bibitem{Bietenholz:2011qq} 
  W.~Bietenholz {\it et al.},
  Phys.\ Rev.\ D {\bf 84}, 054509 (2011)
  doi:10.1103/PhysRevD.84.054509
  [arXiv:1102.5300 [hep-lat]].

\bibitem{Alexandrou:2014sha} 
  C.~Alexandrou, V.~Drach, K.~Jansen, C.~Kallidonis and G.~Koutsou,
  Phys.\ Rev.\ D {\bf 90}, no. 7, 074501 (2014)
  doi:10.1103/PhysRevD.90.074501
  [arXiv:1406.4310 [hep-lat]].

\bibitem{Christ:2010dd} 
  N.~H.~Christ {\it et al.},
  Phys.\ Rev.\ Lett.\  {\bf 105}, 241601 (2010)
  doi:10.1103/PhysRevLett.105.241601
  [arXiv:1002.2999 [hep-lat]].

\bibitem{Dudek:2011tt} 
  J.~J.~Dudek, R.~G.~Edwards, B.~Joo, M.~J.~Peardon, D.~G.~Richards and C.~E.~Thomas,
  Phys.\ Rev.\ D {\bf 83}, 111502 (2011)
  doi:10.1103/PhysRevD.83.111502
  [arXiv:1102.4299 [hep-lat]].


\bibitem{Gregory:2011sg} 
  E.~B.~Gregory {\it et al.} [UKQCD Collaboration],
  Phys.\ Rev.\ D {\bf 86}, 014504 (2012)
  doi:10.1103/PhysRevD.86.014504
  [arXiv:1112.4384 [hep-lat]].

\bibitem{Michael:2013gka} 
  C.~Michael {\it et al.} [ETM Collaboration],
  Phys.\ Rev.\ Lett.\  {\bf 111}, no. 18, 181602 (2013)
  doi:10.1103/PhysRevLett.111.181602
  [arXiv:1310.1207 [hep-lat]].

\bibitem{Bernard:2010fr} 
  C.~Bernard {\it et al.} [Fermilab Lattice and MILC Collaborations],
  Phys.\ Rev.\ D {\bf 83}, 034503 (2011)
  doi:10.1103/PhysRevD.83.034503
  [arXiv:1003.1937 [hep-lat]].

\bibitem{Gregory:2010gm} 
  E.~B.~Gregory {\it et al.},
  Phys.\ Rev.\ D {\bf 83}, 014506 (2011)
  doi:10.1103/PhysRevD.83.014506
  [arXiv:1010.3848 [hep-lat]].


\bibitem{Dowdall:2012ab} 
  R.~J.~Dowdall, C.~T.~H.~Davies, T.~C.~Hammant and R.~R.~Horgan,
  Phys.\ Rev.\ D {\bf 86}, 094510 (2012)
  doi:10.1103/PhysRevD.86.094510
  [arXiv:1207.5149 [hep-lat]].

\bibitem{Mohler:2011ke} 
  D.~Mohler and R.~M.~Woloshyn,
  Phys.\ Rev.\ D {\bf 84}, 054505 (2011)
  doi:10.1103/PhysRevD.84.054505
  [arXiv:1103.5506 [hep-lat]].





\bibitem{'tHooft:1973jz}
  G.~'t Hooft,
  Nucl.\ Phys.\ B {\bf 72}, 461 (1974).
  doi:10.1016/0550-3213(74)90154-0



\bibitem{Witten:1979kh}
  E.~Witten,
  Nucl.\ Phys.\ B {\bf 160}, 57 (1979).
  doi:10.1016/0550-3213(79)90232-3

\bibitem{DeGrand:2016pur} 
  T.~DeGrand and Y.~Liu,
  Phys.\ Rev.\ D {\bf 94}, no. 3, 034506 (2016)
  Erratum: [Phys.\ Rev.\ D {\bf 95}, no. 1, 019902 (2017)]
  doi:10.1103/PhysRevD.95.019902, 10.1103/PhysRevD.94.034506
  [arXiv:1606.01277 [hep-lat]].



\bibitem{Lucini:2012gg}
  B.~Lucini and M.~Panero,
  Phys.\ Rept.\  {\bf 526}, 93 (2013).
  doi:10.1016/j.physrep.2013.01.001
  [arXiv:1210.4997 [hep-th]].

\bibitem{Bali:2013kia}
  G.~S.~Bali, F.~Bursa, L.~Castagnini, S.~Collins, L.~Del Debbio, B.~Lucini and M.~Panero,
  JHEP {\bf 1306}, 071 (2013).
  doi:10.1007/JHEP06(2013)071
  [arXiv:1304.4437 [hep-lat]].



\bibitem{Donini:2016lwz} 
  A.~Donini, P.~Hernández, C.~Pena and F.~Romero-López,
  Phys.\ Rev.\ D {\bf 94}, no. 11, 114511 (2016)
  doi:10.1103/PhysRevD.94.114511
  [arXiv:1607.03262 [hep-ph]].




\bibitem{Witten:1983tx}
  E.~Witten,
  Nucl.\ Phys.\ B {\bf 223}, 433 (1983).
  doi:10.1016/0550-3213(83)90064-0



\bibitem{Adkins:1983ya}
  G.~S.~Adkins, C.~R.~Nappi and E.~Witten,
  Nucl.\ Phys.\ B {\bf 228}, 552 (1983).
  doi:10.1016/0550-3213(83)90559-X

\bibitem{Gervais:1983wq}
  J.~L.~Gervais and B.~Sakita,
  Phys.\ Rev.\ Lett.\  {\bf 52}, 87 (1984).
  doi:10.1103/PhysRevLett.52.87


\bibitem{Gervais:1984rc}
  J.~L.~Gervais and B.~Sakita,
  Phys.\ Rev.\ D {\bf 30}, 1795 (1984).
  doi:10.1103/PhysRevD.30.1795




\bibitem{Jenkins:1993zu}
  E.~E.~Jenkins,
  Phys.\ Lett.\ B {\bf 315}, 441 (1993).
  doi:10.1016/0370-2693(93)91638-4
  [hep-ph/9307244].


\bibitem{Dashen:1993jt}
  R.~F.~Dashen, E.~E.~Jenkins and A.~V.~Manohar,
  Phys.\ Rev.\ D {\bf 49}, 4713 (1994)
  Erratum: [Phys.\ Rev.\ D {\bf 51}, 2489 (1995)].
  doi:10.1103/PhysRevD.51.2489, 10.1103/PhysRevD.49.4713
  [hep-ph/9310379].


\bibitem{Dashen:1994qi}
  R.~F.~Dashen, E.~E.~Jenkins and A.~V.~Manohar,
  Phys.\ Rev.\ D {\bf 51}, 3697 (1995).
  doi:10.1103/PhysRevD.51.3697
  [hep-ph/9411234].



\bibitem{Jenkins:1995td}
  E.~E.~Jenkins and R.~F.~Lebed,
  Phys.\ Rev.\ D {\bf 52}, 282 (1995).
  doi:10.1103/PhysRevD.52.282
  [hep-ph/9502227].

\bibitem{Dai:1995zg}
  J.~Dai, R.~F.~Dashen, E.~E.~Jenkins and A.~V.~Manohar,
  Phys.\ Rev.\ D {\bf 53}, 273 (1996).
  doi:10.1103/PhysRevD.53.273
  [hep-ph/9506273].

\bibitem{Cherman:2012eg}
  A.~Cherman, T.~D.~Cohen and R.~F.~Lebed,
  Phys.\ Rev.\ D {\bf 86}, 016002 (2012).
  doi:10.1103/PhysRevD.86.016002
  [arXiv:1205.1009 [hep-ph]].



\bibitem{Jenkins:2009wv}
  E.~E.~Jenkins, A.~V.~Manohar, J.~W.~Negele and A.~Walker-Loud,
  Phys.\ Rev.\ D {\bf 81}, 014502 (2010).
  doi:10.1103/PhysRevD.81.014502
  [arXiv:0907.0529 [hep-lat]].



\bibitem{DeGrand:2012hd}
  T.~DeGrand,
  Phys.\ Rev.\ D {\bf 86}, 034508 (2012).
  doi:10.1103/PhysRevD.86.034508
  [arXiv:1205.0235 [hep-lat]].

\bibitem{DeGrand:2013nna}
  T.~DeGrand,
  Phys.\ Rev.\ D {\bf 89}, no. 1, 014506 (2014).
  doi:10.1103/PhysRevD.89.014506
  [arXiv:1308.4114 [hep-lat]].


\bibitem{Cordon:2014sda}
  A.~Calle Cord\'on, T.~DeGrand and J.~L.~Goity,
  Phys.\ Rev.\ D {\bf 90}, no. 1, 014505 (2014).
  doi:10.1103/PhysRevD.90.014505
  [arXiv:1404.2301 [hep-ph]].




\bibitem{DeRujula:1975qlm} 
  A.~De Rujula, H.~Georgi and S.~L.~Glashow,
  Phys.\ Rev.\ D {\bf 12}, 147 (1975).
  doi:10.1103/PhysRevD.12.147



\bibitem{Cirigliano:2019jig} 
  V.~Cirigliano, Z.~Davoudi, T.~Bhattacharya, T.~Izubuchi, P.~E.~Shanahan, S.~Syritsyn and M.~L.~Wagman,
  arXiv:1904.09704 [hep-lat].

\bibitem{Lehner:2019wvv} 
  C.~Lehner {\it et al.},
  arXiv:1904.09479 [hep-lat].


\bibitem{Banks:1981nn} 
  T.~Banks and A.~Zaks,
  Nucl.\ Phys.\ B {\bf 196}, 189 (1982).
  doi:10.1016/0550-3213(82)90035-9

\bibitem{Caswell:1974gg} 
  W.~E.~Caswell,
  Phys.\ Rev.\ Lett.\  {\bf 33}, 244 (1974).
  doi:10.1103/PhysRevLett.33.244












\end{thebibliography}
\end{document}